\documentclass{aa}
\usepackage{graphicx}
\usepackage{amssymb}
\usepackage{natbib}
\def\l{$\lambda$}
\def\mbh{$M_{\rm BH}$\/}
\def\lledd{$L/L_{\rm Edd}$}

\def\Mb{$M_{\rm B}$\/}
\def\feiiq{\rm Fe{\sc ii}$\lambda$4570\/}

\def\msol{M$_\odot$\/}

\def\ltsima{$\; \buildrel < \over \sim \;$}
\def\simlt{\lower.5ex\hbox{\ltsima}}            
\def\gtsima{$\; \buildrel > \over \sim \;$}

\def\simgt{\lower.5ex\hbox{\gtsima}}            

\def\civ{{\sc{Civ}}$\lambda$1549\/}

\def\cm3{cm$^{-3}$\/}
\def\hb{{\sc{H}}$\beta$\/}

\def\hbbc{{\sc{H}}$\beta_{\rm BC}$\/}
\def\hbnc{{\sc{H}}$\beta_{\rm NC}$\/}
\def\mgii{{Mg\sc{ii}}$\lambda$2800\/}

\def\oiiiopt{{\sc{[Oiii]}}$\lambda\lambda$4959,5007\/}
\def\o4363{{\sc{[Oiii]}}$\lambda$4363\/}

\def\hei{{\sc{Hei}}$\lambda$5016\/}

\def\feiiopt{{Fe \sc{ii}}$_{\rm opt}$\/}
\def\feii{{Fe\sc{ii}}\/}
\def\fe{{\sc{Fe}}\/}

\def\vr{{$v_{\mathrm r}$}}
\def\heii{{\sc{Heii}}$\lambda$4686\/}
\def\fe76087{{\sc [Fe vii]}$\lambda$6087\/}
\def\oiii{{\sc [Oiii]}$\lambda$5007}

\def\kms{km~s$^{-1}$}
\def\rk{$R_{\rm K}$\/}
\def\ergss{ergs s$^{-1}$\/}

\def\hi{H{\sc i}\/}
\def\rk{{$R{\rm _K}$}\/}
\def\heii{{{\sc H}e{\sc ii}}$\lambda$4686\/}
\def\hei{{{\sc H}e{\sc i}}\/}
\def\rb{$r_{\rm BLR}$\/}

\def\mbulge{$M_{\mathrm{bulge}}$}
\begin{document}

\title{VLT/ISAAC Spectra of the H$\beta$ Region in Intermediate-Redshift
Quasars\\ II. Black Hole Mass and Eddington Ratio\thanks{Based on observations collected at the European Southern
Observatory, Chile. Proposal ref.: ESO 072.B-0338(A)}}
   \author{J. W. Sulentic\inst{1},
           P. Repetto\inst{2,3},
           G. M. Stirpe \inst{4},
            P. Marziani \inst{2},
            D. Dultzin-Hacyan\inst{3},
                \and
            M. Calvani\inst{2}}

   \offprints{J. W. Sulentic}

\institute{                Department of Physics and Astronomy, University of
              Alabama, Tuscaloosa, AL 35487, USA\\
              \email{giacomo@merlot.astr.ua.edu}
\and          Osservatorio Astronomico di Padova, INAF,
              Vicolo dell' Osservatorio 5, 35122 Padova, Italy\\
              \email{marziani@pd.astro.it; calvani@pd.astro.it}
\and
              Instituto de Astronom\'\i a, UNAM, Aptdo. Postal 70-264, M\'exico, D.  F.
                04510, M\'exico \email{prepetto@astroscu.unam.mx,deborah@astroscu.unam.mx}
\and
               Osservatorio Astronomico di Bologna, INAF,
               Via Ranzani 1, 40127   Bologna, Italy\\
               \email{stirpe@bo.astro.it}
             }

\authorrunning{Sulentic et al. }
\titlerunning{\hb\ in Intermediate $z$\ Quasars: \mbh\ and \lledd}

\date{Received / Accepted  }

\abstract{We derive black hole masses for a sample of about 300
AGNs in the redshift range $0 < z < 2.5$. We use the same virial
velocity measure (FWHM \hbbc) for all sources which represents a
significant improvement over previous studies. We review methods
and caveats for determining AGN black hole masses via the virial
assumption for motions in the gas producing low ionization broad
emission lines. We derive a corrected FWHM(\hbbc) measure for the
broad component of \hb\ that better estimates the virialized line
emitting component by comparing our FWHM measures with a sample of
reverberated sources with \hb\ radial velocity dispersion
measures. We also consider the FWHM of the \feiiq\ blend as a
potential alternative velocity estimator. We find  a range of
black hole mass between $\log $\mbh\ $\sim$ 6.0--10.0, where \mbh\
is in solar masses. Estimates using corrected FWHM(\hb), as well
as FWHM(\feii) measures, reduce the number of sources with $\log$
\mbh\ $>$ 9.5 and suggest that extremely large \mbh\ values
($\log$ \mbh $\ga$ 10) may not be realistic. Derived \lledd\
values show no evidence for a significant population of
super-Eddington radiators especially after correction is made for
sources with extreme orientation to our line of sight. Sources
with  FWHM(\hbbc) $\la$ 4000 \kms\ show systematically higher
\lledd\ and lower \mbh\ values than broader lined  AGNs (including
almost all radio-loud sources). }

\maketitle

\newpage\newpage
\section{Introduction \label{intro}}

Gravitational accretion onto supermassive black holes is generally
accepted as the ultimate energy source of Active Galactic Nuclei
(AGNs). The last decade has seen a major effort to derive
reasonable estimates of black hole masses (\mbh) by assuming
virialized motions in the broad line emitting gas:

\begin{equation}
M_{\rm BH} =   \frac{v^2 r_{\rm BLR}}{G} \label{eqn:virial}
\end{equation}

where $ v $\ is the velocity dispersion of the emitting gas at
distance \rb. The velocity dispersion can be written as $ v = f
{\rm FWHM} $\ where FWHM is the full width half maximum measured
for a suitable emission line. The factor $ f $\ depends on the
geometry and details of the kinematics
\citep{krolik01,mcluredunlop01,onkenetal04}. The usefulness of the
virial assumption for \mbh\ determination is best exploited using
reverberation-mapping studies, especially for the Balmer lines
\citep{koratkargaskell91,petersonwandel99,wandeletal99,
kaspietal00,petersonetal04}. \citet{kaspietal00,kaspietal05}
derive a relation between Broad Line Region (BLR) distance $r_{\rm
BLR}$\ and continuum luminosity

\begin{equation} r_{\rm BLR} \propto (\lambda L_\lambda )^{\alpha} \label{eqn:kaspi}
\end{equation}
where the exponent $\alpha $\ is constrained between 0.5 and 1.0, and is most likely  $\approx$ 0.6--0.7 for broad
\hb\ (\hbbc) and optical continuum luminosity.

There are caveats associated with this method. The virial
assumption is not likely to be generally valid for the emission
line gas in AGNs. It has been known for several decades that
different emission lines in a source can show different width and
profile shape \citep[e.g.,][]{derobertis85,sulentic89}. The virial
assumption implies that the velocity dispersion will steadily
decrease with distance from the central black hole $\propto
r^{-\frac{1}{2}}$. The time lag between continuum fluctuations and
corresponding emission line responses will therefore
anti-correlate with line width. This trend  has been confirmed in
a few objects \citep{petersonwandel00}.

Observations however show that profile width and shape depend on
the ionization potential. The strongest high-ionization lines
(HILs; e.g. \civ) often display blueward asymmetric profiles, or
even centroid blueshifts (up to several 1000 \kms) with respect to
the best estimates of the rest frame of the source \citep[e.g.,
][]{gaskell82, marzianietal96,richardsetal02,bachevetal04}.
Blueshifts like those observed for \civ\ are indicative of
obscuration and radial motions which invalidate the virial
assumption.

Low ionization lines (LILs) like \hb\ are the best candidates for
emission arising from a virialized medium. All or much of the
Balmer (and \feii) line emission is thought to arise from an
accretion disk or a flattened cloud distribution near the disk
\citep[e.g.,][]{collinsouffrinetal88,marzianietal96}. A major
caveat, especially for the Balmer lines, involves the possibility
that: (1) there are two or more emission components in a line, and
(2) only one of them may arise in a region where the virial
assumption holds. The Balmer lines do not usually show very large
line shifts (i.e., shift $\leq$ FWHM) although profiles can be
very asymmetric. Both red and blue asymmetries are observed for
\hbbc\ which is the most studied line because it is relatively
unblended and is observable with optical spectrometers up to $z
\approx 1$\ \citep{osterbrockshuder82,sulentic89,stirpe90,
sulenticetal90,corbin91, corbinfrancis94}.

The most ambiguous sources from the point of view of \mbh\
determination show FWHM(\hbbc) $\ga$ 4000 \kms\ \citep[Population
B, following][]{sulenticetal00b} and redshifted profiles and/or
red asymmetries. Not all parts of the H$\beta$\ profile respond to
continuum changes in the same way implying that some of the line
emitting gas may be optically thin or, less likely, is not exposed
to the variable \hi\ ionizing radiation (we will return to this
issue in \S \ref{improvemass}). The broad line profile in Pop. B
sources may be due to two distinct emitting regions: (1) an
optically thick classical BLR and (2) a broader and redshifted
very broad component that may be optically thin or marginally
optically thick to the Lyman continuum, originating in a distinct
Very Broad Line Region (VBLR)
\citep{marzianisulentic93,shieldsetal95,sulenticetal00c}. The
redshift of the VBLR component raises doubts that it arises from
virialized gas. A strong BLR response to continuum fluctuations,
coupled with a weak or absent response of the VBLR component, can
lead to an overestimate of FWHM for the virialized BLR component
resulting in an overestimate of \mbh\
\citep{wandeletal99,kaspietal00,vestergaard02}.

Sources with FWHM(\hbbc) $\la$ 4000 \kms\ \citep[Population
A,][]{sulenticetal00b} should provide more reliable \mbh\
estimates. The \hbbc\ profile is usually well fit with  a
symmetric  function \citep{veroncettyetal01,sulenticetal02} and
the BLR emission is thought to arise from a Keplerian disk. The
most unreliable Pop. A sources, in a disk emission scenario,
should be those observed near face-on where the rotational (i.e.
virial) contribution to FWHM(\hbbc)\ is minimal. At least some of
the face-on sources may be identified as the so-called ``blue
outliers" which show a weak and significantly blueshifted
\oiiiopt\ lines
\citep{zamanovetal02,marzianietal03b,aokietal05,boroson05}. The
inferences in this, and the preceding paragraph have emerged from
the Eigenvector 1 (E1) scenario that we have pursued for the past
5+ years \citep[][following
\citet{borosongreen92}]{sulenticetal00a,sulenticetal00b}. The
results reported in this paper appear to confirm them.

There are additional caveats connected  with using FWHM(\hbbc) for
\mbh\ derivations. Line asymmetries can affect \mbh\ estimates for
both Pop. A and B sources. The \vr\ displacement of the line
centroid at fractional intensities (at half maximum,
$c(\frac{1}{2}$)) can give useful information about the
uncertainty of FWHM(\hbbc) measurements. The average displacement
value is a few hundred \kms\ with median $c(\frac{1}{2}$) $\approx
400$ \kms\ \citep{marzianietal03a}. This implies that deviations
from a symmetric profile can affect \mbh\ estimates by $\approx$
40 \%. Uncertainties are also introduced by: (1) contamination
from overlapping/nearby lines such as \feii, \heii, \ and
\oiiiopt\
\citep{osterbrockshuder82,derobertis85,joly88,jacksonbrowne91},
(2) FWHM measures based on single-epoch observations, (3) low
$S/N$\ spectra  and (4) spectra without \hb\ narrow component
(\hbnc) subtraction. The \rb\ - $L_\lambda$\ relation is also not
free from uncertainties. Reverberation mapping-based \mbh\
determinations are certainly affected by the non-negligible radial
extent of the optically thick BLR. The derived \rb\ is  not a very
well defined quantity and $\alpha$\ is also somewhat uncertain
because of the intrinsic scatter in the correlation. Finally,
reverberation data does not exist for high luminosity/redshift
quasars requiring an extrapolation of the \rb\ - $L_\lambda$\
relation  in order to estimate \mbh\ for these sources.

Uncertainties for \mbh\ derivations using single profile
observation of \hbbc\ are estimated to be a factor of 2--3 (at a 1
$\sigma$\ confidence level), but may be as low as $30$\%\ if the
velocity dispersion of the variable part of \hbbc\ profile is
employed as a virial estimator \citep{petersonetal04}. \mbh\
estimates based on the virial relation retain a statistical
validity considering that AGNs span a 5dex range in \mbh\ (10$^5$
\msol $\la$ \mbh\ $\la 10^{10}$ \msol) and that the relation has
now been applied to large samples of objects  \citep[$ \sim$
10$^3$;][]{mcluredunlop04}.

With these considerations in mind, and supported by previous
results, we use the virial relation to compute \mbh\ for $\approx$
280 AGNs in our E1 sample of low redshift (0 $\la z \la 1$)
sources \citep{marzianietal03b} supplemented with 25 intermediate
redshift/high luminosity (1 $\la z \la 2.5$) quasars. We derive
\mbh\ and \lledd\ values in several ways. New VLT-ISAAC data are
presented for 9 sources (\S \ref{obs}) which supplement data
already published for sixteen quasars \citep{sulenticetal04}. Line
measures are presented  in \S \ref{results}. We use \hbbc\ and
\feiiq\ line widths in a consistent way over the redshift range $0
\la z \la 2.5$  (a range of 10$^5$\ in luminosity). We compute
black hole mass \mbh, Eddington ratio \lledd\ (\S \ref{mass}) and
we discuss how mass determinations might be improved (\S
\ref{improvemass}) so that the evolution of \mbh\ and \lledd\ with
redshift can be considered (\S \ref{evolution} and \S
\ref{discussion}).

\section{Observations and Data Reduction \label{obs}}

New intermediate redshift data were obtained between 10/2003-03/2004 in service mode with the infrared spectrometer
ISAAC mounted on VLT1 (ANTU) at the European Southern Observatory. Each spectrum corresponds to a wavelength range (IR
windows sZ, J, sH) that covers the region of redshifted \hbbc\ and \feiiq\ or Fe{\sc ii}$\lambda$5130\ at least in
part. Reduction of quasar spectra and standard stars followed exactly the same procedures described in
\citet{sulenticetal04}. Wavelength calibration yielded rms residuals of 0.4, 0.6 and 0.9~\AA\ in the sZ, J and sH
windows, respectively. Absolute flux scales of the spectra will be inaccurate because atmospheric seeing almost always
exceeded the slit width ($\approx$ 0''.6) resulting in significant light loss.

Table \ref{tab:obs} summarizes the new observations and the basic format is given below the table. All sources come
from the Hamburg-ESO  (HE) quasar survey, which is a flux limited (with limiting $m_{\mathrm B} \approx$ 17.5),
color-selected survey \citep{wisotzkietal00}.   Column 2 of Table \ref{tab:obs} lists the blue apparent magnitudes
from the HE survey papers \citep{reimersetal96,wisotzkietal00} while  Col. 3 lists the source redshift $z$\ computed
as described in section \ref{red}.  The brightest sources of the HE at intermediate redshift were preferentially
selected. Col. 4 indicates whether \oiii\ was used to compute $z$\ as indicative of the source rest frame. The
absolute magnitude \Mb\ reported in Col. 5 was computed by assuming $ H_0 = 70$ \kms\  Mpc$^{-1}$, and relative energy
density  $\Omega_{\mathrm M} = 0.3$\ and $\Omega_{\mathrm \Lambda} = 0.7$. The $k$\ correction was computed for a
spectral index $ a = 0.6 $ ($ S \propto \nu^{-a} $). Col. 6 gives the ratio of log specific fluxes at 6 cm and 4400
\AA\ ($ \log R_ {\rm k}$). In most cases only NRAO VLA Sky Survey (NVSS) upper limits are available. Columns 7-12 give details of the
observations explained in the footnotes. The continuum $S/N$\ estimate given in Col. 12 was measured using a small
region of the spectrum that was as flat and free of line emission as possible.

\section{Data Analysis \label{red}}

\subsection{Redshift Determination and Rest Frame Corrections}

\begin{figure*}
\includegraphics[width=19.6cm, height=21.6cm, angle=0]{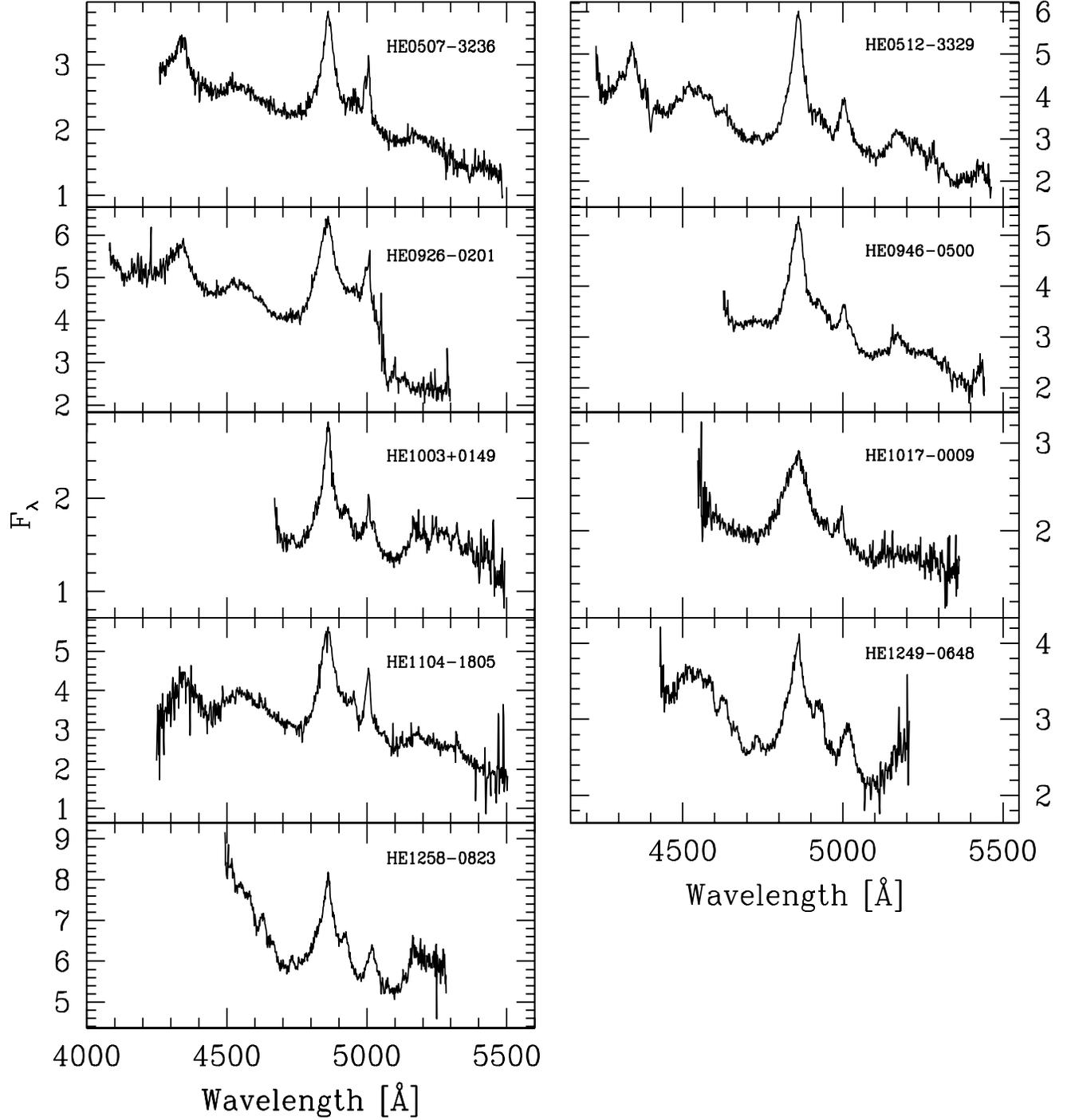}
\caption[originspectr.ps]{Calibrated VLT-ISAAC spectra for 9 new
intermediate-redshift quasars. Absciss\ae\ are rest-frame
wavelength in \AA, ordinates the rest-frame specific flux in units
of 10$^{-15}$ \ergss\ cm$^{-1}$ \AA$^{-1}$.} \label{fig:spectra}
\end{figure*}

Small offsets are present in the wavelength calibration, because the arc lamp frames were obtained in daytime, and
therefore usually after grism movement. A correction for these shifts was obtained by measuring the centroids of 2--3
OH sky lines against the arc calibration and calculating the average difference, which reached at most 6.5~\AA\ or 2.5
pixels in either direction.  Rest frame determination for the 9 new sources was usually estimated from the \hb\ peak
redshift (assumed rest frame $\lambda$ = 4861.33 \AA). \oiii\ (assumed rest frame $\lambda$ = 5006.85 \AA) yielded a
consistent measurement in only two sources (where the results were averaged. Two sources show no clear detection of
\oiiiopt\ while the remaining five show a significant disagreement between \oiii\ and \hbnc. In these sources (we call
the extreme examples  of them ``blue outliers") it is not advisable  to use \oiii\ for redshift determination
\citep{zamanovetal02, marzianietal03b, aokietal05,boroson05}. The adopted estimate was used to deredshift the spectra
while the {\tt dopcor IRAF} task applied a $(1+z)^3$ correction to convert observed specific fluxes into rest frame
values. Fig. \ref{fig:spectra} shows the deredshifted spectra.

\subsection{Continuum and \feii\ subtraction}

Our spectral analysis made use of standard {\tt IRAF} tasks with
the first step involving continuum modelling and subtraction.
Using Chebyshev polynomials of 3$^{\rm rd}$ or 2$^{\rm nd}$ order,
a reasonably smooth continuum subtraction was obtained for all
sources. To estimate errors in the continuum assessment introduced
by noise, we also defined a minimum and a maximum continuum.
Continuum fluxes were chosen at about --3$ \sigma $ (minimum) and
+3$ \sigma $ (maximum), where $\sigma$ is the noise standard
deviation from the most likely continuum choice. Errors in
continuum placement defined by difference between the extreme
continua and the most probable one were then propagated according
to standard error theory. The results of this procedure are
consistent with   continuum fits employing  very simple models
\citep{malkansargent82,shangetal04}. We assumed that the continuum
underlying the \hb\ spectral regions is due to two components:
either a blackbody of temperature 25000$^\circ$ K or a power law
of slope $ b = 0.7 $ ($f_\nu \propto \nu^{-b}$; assumed to be
valid only locally around \hb). In 5 sources the sole black body
component produces a good fit; in 2 the blackbody component is
dominant, and only in the remaining 2 cases the power-law alone
can provide a good fit. This method has some limitations due to
the small spectral bandwidth covered by our spectra, to the
relative strength of \feii\ and to internal reddening effects.
Since we did not attempt to change the blackbody temperature nor
the slope of the power-law, we adopt the empirical continuum which
is visually more accurate.

\begin{figure*}
  \includegraphics[width=19cm, height=19cm, angle=0]{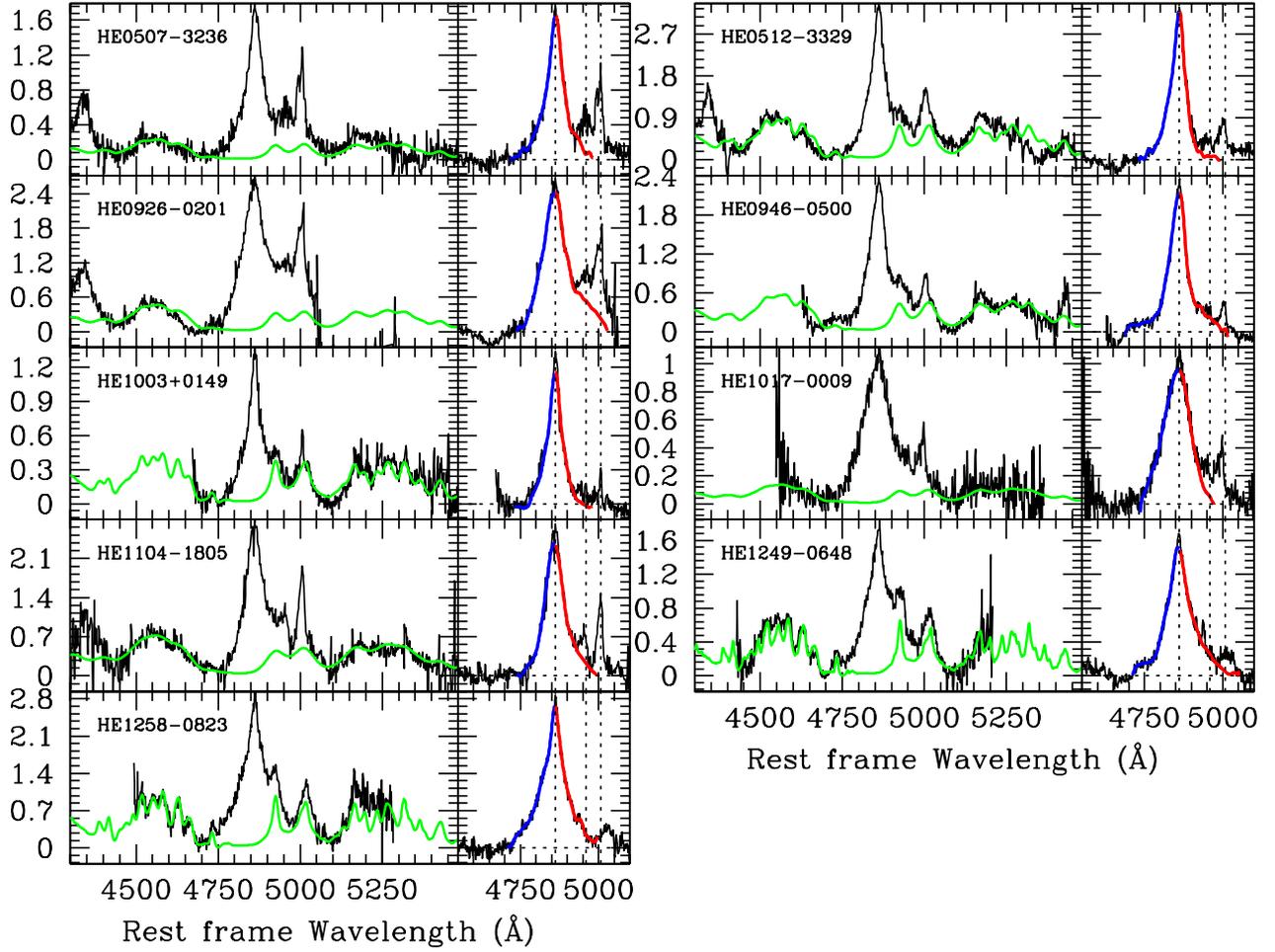}
\caption[hbetaplots.ps]{Spectral Atlas for the new intermediate-redshift quasars. Left-hand panels show the
continuum-subtracted \hb\ spectral region. Axes are the same as Fig. \ref{fig:spectra}. The best-fit \feii\ emission
model is traced as a thin (green) line. Right-hand panels show the enlarged \hb\ profiles after continuum and \feii\
subtraction. The (blue and red) thick line shows a spline fit of the short and long wavelength sides of the  \hbbc\
profile, respectively.}
   \label{fig:hbeta}
   \end{figure*}

The emission blends of \feii\ were subtracted using the template method based upon the spectrum of I Zw 1
\citep{borosongreen92,marzianietal03a}. The strongest \feii\ blends fall in the wavelenghth ranges 4450--4600 \AA\
(blue blend: \feiiq) and 5200--5600 \AA\ (red blend: Fe{\sc ii}$\lambda$5130). The method includes the following
steps: (1) the template intensity is scaled to roughly match the observed spectrum, (2) a Gaussian broadening factor
is estimated from FWHM \hbbc\, (3) a careful estimate of minimum and maximum plausible broadening factor is made to
set a reliable $\pm 3\sigma$\ uncertainty and (4) the template intensity is adjusted as necessary  after broadening.
The resultant template was subtracted from the continuum-subtracted spectra. The blue side of the spectrum including
\feiiq\ is missing, or not fully covered, in several sources (HE 0946--0500, HE 1003+0149, HE 1017--0009, HE
1249--0648 and HE 1258--0823). In these cases the best template fit was achieved for the red blend and a fixed ratio
between the red and blue blends was assumed to reproduce \feiiq.  Fig. \ref{fig:hbeta} shows the estimated \feii\
emission (green lines). \feii\ was subtracted before continuum fitting in sources with strong \feii\ emission.

The \oiiiopt\ lines were measured  after  \feii\ subtraction and
taking into account the following considerations: (1) the flux
ratio between \oiii\ and {\sc [Oiii]}$\lambda$4959 should be
$\approx$3, (2) both lines should show identical profiles and (3)
any \hbbc\ emission underlying \oiiiopt\ is expected to have a
smooth shape. Sources HE~1249--0648 and HE~1258--0823 see (Fig.
\ref{fig:hbeta}) show a small bump at $\lambda \approx 5016$ \AA\
which is either an \feii\ subtraction residual or emission from
He{\sc i}$\lambda$5016 (weak redshifted \oiii\ would be almost
unprecedented). After subtraction of \hbnc\ \citep[following
][]{marzianietal03a}, the \hbbc\ profile was fit with a high order
spline function ({\tt IRAF} task {\tt SFIT}). This procedure does
not yield a model fit but only an empirical fit that smooths the
noise and reproduces the main features and inflections in \hbbc.

\section{Immediate Results \label{results}}

\subsection{Line Measurements and Uncertainties}

\paragraph{Line Fluxes and Equivalent Widths}

Table \ref{tab:profs} gives line measurements for the new VLT
spectra with the  basic format given in the footnote. Cols. 3 and
4 give equivalent width (EW) measures for \hbbc\ and \feiiq\
respectively. We evaluated uncertainties associated with the
continuum level (derived from the minimum and maximum reasonable
continuum estimates) and line flux errors estimated from the
$S/N$. These estimates were combined quadratically to obtain
uncertainties for EW measures. A similar procedure was applied to
obtain uncertainty estimates for \hbnc\ and \oiiiopt.  The EW
uncertainty for \hbnc\ was derived from the estimated maximum and
minimum possible  \hbnc\ component in the \hb\ line. The relative
error of the \hbnc\ flux can be large (see Table \ref{tab:profs})
and in some cases an \hbnc\ component may not be present.

\paragraph{\feiiopt\ width}

\feii\ emission is heavily blended so that widths of individual
lines must be obtained from the best broadening parameter that was
used for the template fit. This requires that we assume a constant
width for all \feii\ lines which so far appears to be reasonable.
FWHM(\feiiq)\ values derived from the template broadening factor
are reported in Col. 5 of Table \ref{tab:profs}. Uncertainty
estimates for FWHM(\feiiq)\ were obtained by increasing/decreasing
the broadening factor until we could detect significant changes in
the best fit. Simulated data reveal that it is possible to
estimate the \feiiq\ width up to FWHM(\feiiq) $\approx$ 6000 \kms.
Due to the very large uncertainty of \feiiq\ width determination
for individual sources, \mbh\ estimates based on  FWHM(\feiiq)
measures are used mainly for confirmatory purposes of statistical
trends detected with \hb.

\paragraph{\hbbc\ Line Profiles}

Measurements of FWHM(\hb) together with other important line parameters like asymmetry index, kurtosis and line
centroid at various fractional intensities were derived using a {\tt FORTRAN} program developed for that purpose.
These parameters are the same as defined in several previous papers
\citep{marzianietal96,marzianietal03a,sulenticetal04} and are given in Table \ref{tab:profs}. Each line measure is
followed in the next Col. by its appropriate uncertainty. Cols. 2 and 4 give the Full Width at Zero Intensity (FWZI)
and FWHM. Col. 6 gives  the asymmetry index (AI) as defined in \citet{sulenticetal04}. Col. 8 lists kurtosis values
while the remaining part of Table \ref{tab:profs} lists measures of the \hbbc\ centroid at various fractional
intensities. (in \kms). All uncertainties  represent the $2 \sigma$\ confidence level.

The dichotomy in \hbbc\ profile shape (and many other properties) between Population A and B \citep{sulenticetal02} is
seen in the new source measures and in the rest of our higher redshift sample \citep{sulenticetal04}. Redward
asymmetries (A.I. $\ga$ 0.2) are most often found in Pop. B sources. They also show \hbbc\ profiles that are best fit
with Gaussian functions, and some profile appear composite. A few sources appear to deviate from the trend found in
previous work that FWHM of \hbbc\ and the \feii\ lines are very similar \citep{marzianietal03a,marzianietal03c}.
HE~1249-0648 and HE~1258$-$0823 show FWHM(\hbbc) $\gg$ FWHM(\feiiq). While FWHM(\feiiq) is subject to large
uncertainty, the difference is confirmed by careful reinspection of these spectra. This condition is seen in only
2/215 sources in the \citet{marzianietal03a} sample. Both (IRAS 07598+6508 and Mkn 235) are BAL QSOs which are also
FIR ultra-luminous \citep{sulenticetal06}. In addition to a \civ\ BAL with high terminal velocity, these objects have
 a  strong, blueshifted \civ\ emission line component. Blueshifted Balmer emission is  probably associated with the
high-ionization gas emitting \civ. The absence of any detectable \oiiiopt\ emission, along with a possible \hei\
feature at $\lambda$5016 \AA\ support the possibility that HE~1249-0648 and HE~1258$-$0823 could be BAL QSOs similar
to IRAS 07598+6508 and Mkn 235. Further support comes from an inspection of the HE survey spectra of both objects
which indeed show blueshifted broad absorption in the \mgii\ line.

\section{Black Hole Mass and Eddington Ratio \label{mass}}

\subsection{Basic Equations}

One can write the velocity dispersion $v$\ in Eq. \ref{eqn:virial}
as $ v \approx  \sqrt{3}/2 $\  FWHM(\hbbc) in the case of randomly
oriented velocities projected along the line-of-sight. The
expression for black hole mass is then:
\begin{equation}
M_{\rm BH} \approx \frac{3}{4} \frac{r_{\rm BLR} {\rm FWHM(\rm
H\beta_{\rm BC})}^2}{G}
\end{equation}

In the absence of reverberation data one must rely on the
correlation between reverberation radius and source luminosity $
r_{\rm BLR} \propto \left(L_{5100}\right)^{\alpha}$, where $
L_{5100} $\ is the specific luminosity at $\lambda \approx $ 5100
\AA\ \citep{kaspietal00,kaspietal05}. This formula relates BLR
distance from the central continuum source and specific luminosity
(\ergss \AA$^{-1}$) near 5100 \AA. Following \citet{kaspietal05}
one can write:

\begin{equation}\label{eqn:fe}
r_{\mathrm BLR} \approx 0.697 \cdot  10^{17} \cdot \left[\frac{
\lambda L\rm _{\lambda} \left(5100 \AA\right)}{10^{44} \,\rm erg
\, s^{-1}}\right]^{0.67} {\rm cm}
\end{equation}

\noindent   We can make two estimates of \mbh\ using these
relationships assuming two different values for $\lambda
L_\lambda$. The first method derives \Mb\ from tabulated values of
$V$\ or $B$ apparent magnitudes:

\begin{equation}\label{eqn:te}
\lambda L_{\lambda}\left(5100{\rm \AA}\right) \approx 3.14 \cdot
10^{35 - 0.4 (M_B)} ~\mathrm{ergs ~s^{-1}}
\end{equation}

A second possibility involves estimating the specific luminosity near 5100 \AA\ directly from flux measures using the
adopted continuum fits for our spectra. More explicitly $\lambda L_\lambda$\ is computed as follows:

\begin{equation}
\lambda L_{\lambda}\left(5100 {\rm \AA}\right) = 4 \pi d_{\rm P}^2
\lambda f_\lambda (5100 \mathrm \AA)~ \rm ergs ~s^{-1},
\end{equation}

where $d_{\mathrm P}$\ is the redshift derived distance and $f_\lambda$\ is the specific flux in the rest frame at
5100 \AA\ (after correction for Galactic extinction $A_{\mathrm B}$\ at the observed wavelength). The second choice
has the advantage that the continuum $f_\lambda$\ and FWHM(\hbbc) are measured from the same spectrum. Expected light
losses are estimated to be $\approx$ 35 \%\ of the quasar flux with average Paranal seeing (0.''6). We apply this
average correction to the observed flux before computing \mbh.

Substituting the expression for \rb\ in the mass formula one
obtains the following relation:

\begin{equation}\label{eqn:foe}
M_{\rm BH} \approx 5.48 \cdot 10^6  \left[\frac{\lambda
L_{\lambda} \left(5100\rm \AA\right)}{10^{44} \,\rm erg \,
s^{-1}}\right]^{0.67} \left(\frac{\rm FWHM(H\beta_{BC})}{1000 \rm
km \, s^{-1}}\right)^2
\end{equation}


\noindent where \mbh\ is in solar mass units (\msol).

Fig. \ref{fig:massz} shows the distribution of \mbh\ as a function of $z$. \mbh\ estimates are based on the
 FWHM(\hbbc) measures reported in Table \ref{tab:meas}.  All sources with $z \ga 0.8$ are from
the VLT sample. We do not show \mbh\ values derived  from \Mb\ as
in \citet{marzianietal03b} for clarity, since they basically
confirm the same trends obtained from the specific fluxes.


\subsection{Improved \mbh\ Estimators? \label{improvemass}}

As briefly summarized in \S \ref{intro}, there is some evidence
suggesting that the virial assumption is reasonable for LILs in a
significant fraction of quasars (Population A; about 50-60\% of
low $z$\ quasars). Our VLT spectra confirm that any dependence of
FWHM(\hbbc) on source luminosity is weak \citep{sulenticetal04}.
Table \ref{tab:profs} suggests that low redshift trends for
asymmetries and line shifts are preserved in the intermediate $z$\
sample. It is therefore not certain that FWHM(\hbbc) is a valid
estimator of the virial velocity for all sources even after proper
\hbnc, \oiiiopt, \feii\ subtraction.

A more physical approach to \mbh\ estimation uses  the FWHM  of
the variable part of the \hbbc\ profile \citep{petersonetal04}.
Fig. \ref{fig:peterson} plots FWHM$_{\mathrm{rms}}$\ values from
\citet{petersonetal04} versus our FWHM(\hbbc) measures
\citep{marzianietal03a} for all sources in common. Sources with
FWHM $\la$ 4000 \kms\ show a correlation while the situation is
less clear for sources with FWHM$\ga$ 4000 \kms. We recall that
this value indicates the nominal population A-B boundary
\citep{sulenticetal00b} that emerged in our E1 studies. A
least-square best fit analysis yields a ``corrected" FWHM(\hbbc)
estimate: FWHM$^{\rm corr}$(\hbbc)$ \approx -710 (\pm 800) + 1.13
(\pm 0.28) ~ $ FWHM(\hbbc) for FWHM $\la$ 4000 \kms\ and
FWHM$^{\rm corr}$(\hbbc)$ \approx 650(\pm 1000) + 0.79(\pm 0.14) ~
$FWHM(\hbbc) for FWHM $\ga$ 4000 \kms. The slope depends somewhat
on the fitting method (a robust fit yields 0.94 and 0.72) but is
always less steep for FWHM $\ga$ 4000 \kms. An immediate
implication is that optically thick BLR gas responding to
continuum changes shows a velocity dispersion correlated with --
but slightly lower than that of the integrated profile at all
FWHM.

A break in the linear fit at 4000 \kms\ is consistent  with
several previous findings: (a) mean and possibly systematic \hbbc\
profile differences between Pop. A and B sources
\citep{sulenticetal02}; (b) the lack of strong profile changes in
Pop. A sources \citep{giveonetal99}; (c) profile asymmetries
frequently observed in Pop. B sources. They might represent a
distinct redshifted emission component \citep{sulenticetal02}which
may arise in less optically thick gas than the rest of the \hbbc\
profile \citep{sulenticetal00c}. If the redshifted component is
emitted in an innermost VBLR, the FWHM of the whole \hbbc\ profile
is obviously increased over the value due to the line component
that is actually responding, and that is most likely located
farther away from the central continuum source. Even if the virial
assumption holds for the VBLR (but the frequent asymmetries warn
us that this might not be the case), the use of the FWHM from the
whole \hbbc\ profile and of \rb\ from the reverberating part,
yields  an \mbh\ overestimate.  This interpretation of the
correlations in Fig. \ref{fig:peterson} is advanced with caution
because of the small sample size and especially because of the
poor statistics for sources with FWHM $\ga$ 4000 \kms. It is also
possible to produce a meaningful fit with a second-order
polynomial. We apply a tentative correction to the FWHM measures
FWHM$^{\mathrm{corr}}$(\hbbc)\ and  therefore to resultant \mbh\
estimates  using the second-order  fit shown in Fig.
\ref{fig:peterson} which approximates very well the linear trends.

If an optically thin/nonvirialized component is present in the \hbbc\ profile of many sources then we might use as a
virial estimator the FWHM of a line, or lines,  arising in BLR gas but not likely to be present in the VBLR region.
FWHM(\feiiq) is an obvious alternative because there is no evidence for a VBLR emission component in the broad \feii\
blends. It is not strictly correct to use it because the \rb - $L_\lambda$\ relation was deduced for \hb. The most
serious  difficulty lies in  obtaining a reliable FWHM(\feiiq) estimate from the heavily blended \feii\ emission.
Considering the EW and FWHM limits for detection of \feiiopt\ emission \citep{marzianietal03a} we conclude that a
reasonable FWHM measurement is possible for $\approx$120 sources in our low $z$\ spectral atlas. Measurement
uncertainties for FWHM(\feii) will be larger than for FWHM(\hbbc) and are estimated to lie  between 20--50\%.  The
best fit of FWHM(\feiiq) vs. FWHM(\hbbc) (\kms) is consistent with FWHM(\feiiq) $\approx$ FWHM(\hbbc) for FWHM(\hbbc)
$\simlt$ 4000 \kms, while it is FWHM(\feiiq) $\approx$ 0.67 $\cdot$ FWHM(\hbbc) + 820 \kms\ if  FWHM(\hbbc)$\simgt$
4000 \kms.

The relationship between  FWHM(\hbbc) and FWHM (\feii) confirms
that individual \feii\ lines show approximately the same width as
\hbbc\ and as the rms \hbbc\ component implying a common kinematic
environment if FWHM(\hbbc) $\la $ 4000 \kms.  If FWHM(\hbbc) $\ga$
4000 \kms, FWHM(\feii) follows a trend closer to that of the rms
\hbbc\ component. Therefore if  the rms \hbbc\  component arises
from gas in virialized motion, the same can be reasonably assumed
for the whole \feiiq\ emission.

\subsection{\mbh\ and \lledd\ Dependence on Redshift \label{evolution}}

\subsubsection{\mbh}

Fig. \ref{fig:massz} shows the distribution with $z$\ of \mbh\ estimates derived using uncorrected FWHM(\hbbc)
measures. This can be compared with Fig. \ref{fig:masszpet} where we show corresponding distributions of \mbh\ derived
using corrected FWHM$^\mathrm{corr}$(\hbbc)\ measures and using FWHM(\feiiq). FWHM$^\mathrm{corr}$(\hbbc)\ values were
derived from the second-order relation in Fig. \ref{fig:peterson}.


The distribution of data points in the  $\log$ \mbh\ vs. $z$\
plane reflects selection effects intrinsic to any flux limited
sample. If we consider AGNs that are radiating at a given \lledd\
we easily obtain \mbh\ for a given apparent magnitude:

\begin{eqnarray*}
{M_{{\mathrm{BH}},9\odot}} = &   & {3.85 \cdot 10^6 h^{-2}}
{\left( \frac{L}{L_{\mathrm Edd}}\right) }^{-1}  \cdot
10^{-0.4 m_{\mathrm B}} \cdot \\
& &  \cdot \left[ 1.5  \left( 1 - e^{-\frac{z}{6.107 }} \right) +
 \left( 1 - e^{-\frac{z}{1.266}} \right) \right]^2  \cdot (1 + z)^{(1 - \mathrm{a})}. \\ \label{eqn:mass}
\end{eqnarray*}

Appendix \ref{appendix} details how Eqn. \ref{eqn:mass} was derived. If we adopt a limiting magnitude $m_{\mathrm B}
\approx 17.5$\ (appropriate for the Hamburg-ESO survey) and consider a maximum Eddington ratio \lledd $\approx$ 1 we
obtain the minimum \mbh\ detectable as a function of $z$. Limiting \mbh\ curves are shown in Fig. \ref{fig:massz},
Fig. \ref{fig:masszpet}, and Fig. \ref{fig:masszrlq}.

All sources show \mbh\  $\ga 10^6$ \msol. In the range  $1 \la z \la 2.5$\ we begin to find a significant black hole
population in the range $\sim$ $10^{9.5}$ \msol\ $\la$ \mbh\ $\la 10^{10}$ \msol. If we consider \mbh\ values derived
from \feiiq\  as well as from FWHM$^{\mathrm{corr}}$(\hbbc) (Fig. \ref{fig:masszpet}, open and filled symbols
respectively) we get a somewhat different picture. Almost all sources lie below $\approx 5 \cdot 10^{9}$ \msol\ (all
sources with $z \la 1$); only three sources whose \mbh\ has been computed with from FWHM$^{\mathrm{corr}}$(\hbbc) lie
significantly above this limit. Our sample of intermediate $z$\ quasars with compatible quality data is still small so
caution is needed. It should be pointed out that the use of \hbbc$^{\mathrm{corr}}$\ is self-consistent, since no \hb\
surrogate line was used.

The upper panel of Fig. \ref{fig:masszrlq} identifies sources on
the basis of radio-loudness where open and solid symbols denote
radio-loud \citep[RL; see definition in][]{sulenticetal03} and
radio-quiet (RQ) sources respectively. At low $z$\ RQ sources  are
distributed across the full range of \mbh\ up to \mbh $\ga 10^9$
\msol, although RL sources tend to have larger \mbh\  \citep[][and
references therein]{metcalfmagliocchetti06,marzianietal03b}. The
same consideration apply to Pop. A and Pop. B sources (lower panel
of Fig. \ref{fig:masszrlq}), and this is not surprising since most
Pop. A sources are RQ. At intermediate $z$, selection effects
limit the detectable black holes to \mbh $\ga 10^9$ \msol, so that
only the high end of the quasar \mbh\ distribution can be traced.
It is too early to decide whether systematic differences in the
\mbh\ distribution of RQ and RL sources may still exist at
intermediate $z$, at least from the present data.

\subsubsection{\lledd}

Fig. \ref{fig:lmz} shows the distribution of \lledd\ estimates  as a function of $z$.  A preliminary analysis in terms
of \lledd, and including discussion of bolometric luminosity estimation, was presented in \citet{marzianietal03b}.


The existence of apparently super-Eddington radiators in the
low-$z$\ part of Fig. \ref{fig:lmz} involves the most extreme
NLSy1 sources. One caveat about interpreting any of them as
super-Eddington involves the significant uncertainties associated
with these estimates. If we assume a typical uncertainty of
$\pm$50\%\ for the bolometric correction (neglecting beaming or
lensing) and $\pm$10\%\ for the virial estimator (basically the
uncertainty in the FWHM measurement) we find a  $\Delta \log
$\lledd $\approx$ 0.13 (neglecting the scatter in Eqn.
\ref{eqn:kaspi}). A serious source of uncertainty ignored until
now involves source orientation with respect to the line-of-sight.
The virial velocity dispersion that \hbbc\ is assumed to measure
is now widely assumed to involve Keplerian rotation in an
accretion disk or in a flattened distribution near the disk. If
our candidate super-Eddington sources are mostly viewed with
(face-on disk and/or pole-on jet) then \mbh\ can be significantly
underestimated. An orientation correction applied to extreme NLSy1
sources (we call them Pop. A ``blue outliers") interpreted as
face-on, will move them below \lledd\ $\approx$ 1 \citep[][see
arrows in their Figure 12]{marzianietal03b}. At high $z$\ we so
far find no strong evidence for any super-Eddington sources. The
thick solid line in Fig. \ref{fig:lmz} shows the expected \lledd\
detection limit deduced from Eqn. \ref{eqn:mass} for a
flux-limited survey ($m_{\mathrm B} \approx 17.5$) and a quasar
with \mbh\ $\approx 4 \cdot 10^9$ \msol. All quasars in our sample
fall above the minimum detectable \lledd, suggesting that masses
larger than $\approx 4 \cdot 10^9$\msol\ are not strictly
necessary from our data (see \S \ref{largemass} for further
discussion).


A physical basis for the  dichotomy between Pop. A and Pop. B
\citep{sulenticetal00b,marzianietal01,marzianietal03b} is
supported by this analysis in the sense that Fig.
\ref{fig:peterson} shows evidence for a change at about
FWHM(\hbbc) $\approx$ 4000 \kms.  Fig. \ref{fig:lmz} identifies
Pop. A and B sources as filled and open symbols respectively. Here
we apply the luminosity-dependent definition of the Pop. A-B
boundary as defined in Fig. 6 of \citet{sulenticetal04}. It shows
that the Eddington ratio  of Pop. A sources is systematically
larger than that of Pop. B, and that the apparent boundary between
the two populations may increase with redshift.  If we focus on
the intermediate $z$\ quasars then Pop. A sources show an average
\lledd\ $\approx$ 0.78 compared to $0.27$\ for 10 Pop. B sources.
Even these small samples of Pop. A and B sources show
significantly different \lledd\ distributions according to a
Kolmogorov-Smirnov (K-S) test. RL AGNs are systematically low
\lledd\ radiators since they are almost entirely Pop. B sources.

It is interesting to note that with: (1) a limiting magnitude
$m_{\mathrm B} \approx 16.5$, (2) \lledd $\la$ 1 and (3) \mbh\
$\simlt 4 \cdot 10^9$ \msol\ we should detect fewer sources beyond
$z \approx$ 2 (no source below the dot-dashed curve of Fig.
\ref{fig:lmz}). If our assumptions about the quasar bolometric
correction are valid up to that redshift, selection effects may
influence the relative frequency of Pop. A and B sources (the two
Populations have different \lledd\ distributions) rather than the
intrinsic properties of LILs. In this case selection effects on
\lledd\ should strongly influence the so-called ``Baldwin effect"
involving \civ\ and other HILs \citep[][ and references
therein]{bachevetal04} because \civ\ is more prominent in Pop. B,
and Pop. B sources are more easily lost at high $z$.


\section{Discussion \label{discussion}}

The present paper provides \mbh\ estimates that have three
advantages: (1) a consistent data analysis procedure is employed
over the entire redshift  range 0.0 $\la z \la $ ~2.5 by using the
same \mbh\ tracer, \hbbc; (2) $S/N$\ and resolution of the
spectroscopic data sample are high enough to permit a careful
study of the \hbbc\ profile, and (3) the data quality allows
reasonable estimates of FWHM \feii\ in many of the sources.

\subsection{What are the largest black hole masses? \label{largemass}}

\citet{netzer03} discussed several problems with  \mbh\ $\ga
10^{10}$ \msol. If the black hole mass vs. bulge mass (\mbulge)
relation \citep{ferraresemerritt00} is valid at high $z$\ then
\mbh\ $\sim 10^{10}$ \msol\ would imply stellar velocity
dispersion $\sigma_\star \approx 700$\ \kms\
\citep[following][]{gebhardtetal00}  and resultant bulge masses
\mbulge\ $\ga 10^{13}$\msol\ which are not observed at low-$z$\
\citep{mclurejarvis04,netzer03,wang03}. Recent results for the
fundamental plane of elliptical galaxies, and  the  most massive
spheroids at $ z \simlt 0.3$ from Sloan Digital Sky Survey
measures, confirm that $\sigma_\star \simlt 500$\ \kms\
\citep[SDSS,][all galaxies with $\sigma \simgt 500$ \kms\ are
likely due to chance
superposition]{bernardietal03,bernardietal05}.

There are several proposed interpretations of this problem: (1)
the \mbh\ -- \mbulge\ relationship may not strictly hold for all
hosts, (2) the virial assumption is not  applicable, (3) results
are plagued by such large uncertainties -- including the one of
the luminosity index $\alpha$\ -- that very large mass estimates
are not real \citep{vestergaard02}, and (4) some systematic
effects may not have been considered.

\subsubsection{The \mbh\ -- \mbulge\ Relationship}

Nuclei with \mbh\ $\ga  5 \cdot 10^9$ \msol\ are not observed in
galaxies of the local Universe if a direct black hole mass
determination is possible from circum-nuclear kinematics
\citep{marconihunt03}. They are expected to be rare and difficult
to find, considering also that they should be in a dormant or
nearly dormant stage at the present epoch. Integrating the quasar
luminosity function at $z \approx 1.5$\ \citep{boyleetal00}, we
find a comoving density of quasars above the HE limiting magnitude
(which corresponds to \Mb $\approx -27.1$\ for a $k-$correction $a
= 0.6) \sim 3 \cdot 10^{-8}$ Mpc$^{-3}$. This indicates that the
present-day density of the most massive black holes that were once
luminous quasars should be very low, $\sim 2\cdot 10^{-9}$\
Mpc$^{-3}$. Even so, they should be much more frequent than the
very massive spheroids that would host them if the \mbh --
\mbulge\ relationship is valid. To estimate the density of
spheroids with $\sigma_\star \simgt 500$ \kms, one must
extrapolate the $\sigma$\ distribution function provided by
\citet{shethetal03} to unobserved domains in $\sigma$. Integrating
the \citet{shethetal03} function for $\sigma_\star \simgt 500$
\kms, one obtains that the comoving density of all local spheroids
is three orders of magnitude lower ($\sim $ 6 $\cdot$\ 10$^{-13}$
Mpc$^{-3}$) than that of the most massive black holes. A possible
implication is that the \mbh\ -- \mbulge\ relation is not linear
or of universal validity, i.e. some galaxies  host larger black
hole masses  than expected.  This conclusion should remain valid
also at intermediate redshifts. A constant \mbh/\mbulge\ ratio
cannot hold forever if bulge mass grows by secular evolutionary
processes. Evolution goes in the sense that the spheroids may
increase their masses at a later cosmic age \citep[$ z \la
2$;][]{treuetal04,pengetal05} significantly more than their
central black hole: the most massive black hole should be already
``brightly shining'' at $z \approx 2$\ \citep[][and references
therein]{mclureetal05}. A nonlinear \mbh\ -- \mbulge\ relation has
been proposed \citep{laor01} for local hosts, with \mbh\
accounting for only 0.05 \%\ of the bulge mass in low-luminosity
galaxies and 0.5 \%\ in giant ellipticals. In this case, \mbh
$\sim 10^{10}$ \msol\ would imply \mbulge $\sim 10^{12}$ \msol,
which is at the upper end of the spheroid masses measured locally
\citep{marconihunt03}.

The problem can be made to disappear if \mbh $\la 3 \cdot 10^9$
\msol. \citet{bernardietal05} find a maximum $\sigma_\star \approx
500 $ \kms. These local galaxies  (with expected \mbh\ $\approx 3
\cdot 10^9$ \msol) may have been the hosts of the luminous quasars
at $1 \simlt z \simlt 2$ since the number density of spheroids
with $\sigma_\star \simgt 350$ \kms\ is  $ \sim 4 \cdot  10^{-7} $
Mpc$^{-3}$, still somewhat larger than the present-day density of
supermassive black holes that were radiating at \Mb\ $\la -25$ ($
\sim 10^{-7} $ Mpc$^{-3}$).


The curves in Fig. \ref{fig:lmz}, where  \mbh $\sim 4 \cdot 10^9$
\msol\ has been assumed, suggest that there is no need for \mbh\
$\ga 10^{10}$ \msol\ at $z \simlt 2.5$. Similarly  the brightest
HE sources in the redshift range $1 \simlt z \simlt 2 $\
($m_\mathrm B \approx 15.0$\ at $ z \approx 1.7$)  are consistent
with \lledd $\approx$ 1 if \mbh $\sim 4 \cdot~ 10^9$ \msol. In
other words, if \mbh $\sim 10^{10}$ \msol, then a source at \lledd
$\approx$ 1 would be brighter than the brightest quasars observed
in the range $1 \simlt z \simlt 2$\ (if $H_0$\ is close  to the
value assumed in this paper).

\subsubsection{Could A Non-virial Component Yield a Huge \mbh?}

One might envision a non-virial component that  increases with
source luminosity and systematically broadens the \hbbc\ profile.
In our so-called Pop. A sources, the high ionization wind that can
dominate \civ\ emission might reasonably be expected to produce
Balmer line emission  as well.  Such an additional (blueshifted)
component on the \hbbc\ profile would increase FWHM measures
\citep[][HE 1249--0648 and HE 1258--0823 may show this
effect]{sulenticetal06}. One intriguing possibility is that
high-luminosity sources are  extreme Pop. A sources which is
supported by models invoking a radiation pressure driven wind.
Such an effect would  make the correction ($\sim $ 500 \kms)
deduced for low-$z$\ Pop. A  sources  inadequate. At the other
extreme, strong redward asymmetries  observed in Pop. B sources
indicate that the integrated profile may be affected by
gravitational redshift and non-virial motions
\citep{marzianietal03b}. A quantitative analysis of these
suggestions needs careful analysis beyond the scope of this paper,
but in both cases any correction would lower \mbh.

\subsubsection{Statistical Errors}

It  is interesting to consider in more detail the sources that
show \mbh\ $\sim 10^{10}$ \msol\ in our sample. HE 0248--3628 with
the largest estimated \mbh\ is also one of the highest \lledd\
radiators with \lledd $\sim$ 1. It shows an anomalously large (by
a factor $\sim$10) flux with respect to our other VLT sources
which makes it an extremely luminous quasar \citep[\Mb $\la -30$;
see][]{sulenticetal04}. It falls near the Pop. A--B boundary  in
Eigenvector 1 space and is a borderline radio-loud source by our
definition \citep{sulenticetal03}. It disappears as an outlier if
the  \mbh\ estimated from the tabulated \Mb\ is adopted.  HE
2355--4621 is a second source with \mbh$\sim 10^{10}$ \msol. It
behaves like a normal radio-quiet Pop. B source. The \hbbc\
profile shows a prominent redward asymmetry, which is strongly
affecting the width at half maximum. It is also interesting to
consider HE 1104--1805, which is the most luminous source in our
new sample with (\Mb $\approx $ -29.5).  It behaves like an
ordinary Pop. A source and shows \mbh$\approx 5 \cdot 10^{9}$
\msol\ which may be overestimated if the continuum is lens
brightened (Appendix \ref{individual}).

The high-mass wing of the \mbh\ distribution in the redshift range
$1 \simlt z \simlt 2$\ is consistent with the wing of a Gaussian
peaked at $\approx$ 3 $\cdot 10^9$ \msol\  and dispersion $\Delta
\log$ \mbh\ $\approx$ 0.3 if  \mbh\ is computed from
FWHM$^{\mathrm{corr}}$(\hbbc) (4 $\cdot 10^9$ \msol\ if no
correction to \hbbc\ is applied).  The Gaussian dispersion is
consistent with the estimated errors of individual measurements.
$K-S$\ tests do not favor significantly different peak masses or a
much different dispersion. This suggests that sources with  \mbh\
$\ga$ 3 $\cdot 10^9$ \msol\ of our data might be mostly due to
random errors associated to the uncertainty in individual \mbh\
measurements.

\subsubsection{Systematic Effects}

A first systematic effect considers the uncertainty in the index
$\alpha$. Our \mbh\ estimates show the onset of \mbh $\ga 3 \cdot
10^9$ \msol\ at $z \sim$0.8; at $z \ga$0.8 we  observe some of the
most luminous HE quasars.    The high luminosity range of the
\citet{kaspietal05} relationship remains poorly sampled: there are
just 2-3 sources in the range $45 \la \log \lambda L_\lambda \la
46$\ (to which most of our intermediate $z$\ sources belong),
creating a sample bias and making the correlation analysis
intrinsically unstable. In addition, a recent reanalysis of the
\rb -- $\lambda L_\lambda$\ correlation suggests a value of
$\alpha$ as low as 0.5 \citep{vestergaardpeterson06}. If $\alpha$\
is overestimated  by $\Delta\alpha \approx 0.17$\  over a
luminosity range of $\sim 10$, \mbh\ may be overestimated by
$\Delta \log $ \mbh $\approx$ 0.2 in our intermediate $z$\
sources.

Pop. A sources show good evidence that LILs are emitted in a
strongly flattened system, probably an accretion disk or gas
co-planar with the disk. \civ\ in these sources seems to be
dominated by a wind component \citep{bachevetal04}. If these
considerations apply also to Pop. B sources (but it is by no means
clear, given the large FWHM(\hbbc) of sources believed to be
observed pole-on) and if the maximum angle between  disk axis and
line-of-sight is $\approx 45^\circ$, a correction could imply a
factor $\simlt 2$ systematic increase in \mbh. However, these
consideration of orientation effects may not even reopen the
problem of very large \mbh\ values: if our suggestion of a maximum
\mbh $\approx 3 \cdot 10^9$ \msol\ is appropriate, taking into
account the systematic orientation effects would yield a maximum
\mbh $\approx 6 \cdot 10^9$ \msol\ which is still plausible.  The
result that \lledd $\la$ 1 at $z \ga$ 0.8 would be reinforced by
systematic orientation effects. \medskip

Summing up, our data suggest that very large masses $\sim 10^{10}$
\msol\ may be not be real, and may be predominantly due to
statistical errors and  emission line profile broadening that is
in part non-virial. The data presented in this paper are
consistent with \mbh\ not exceeding $3 \cdot 10^9$ \msol\ for our
sources if the correction to FWHM(\hbbc) described earlier is
applied.

\subsection{The Best \mbh\ Estimators}

\begin{figure*}
  \includegraphics[width=14cm, height=14cm, angle=0]{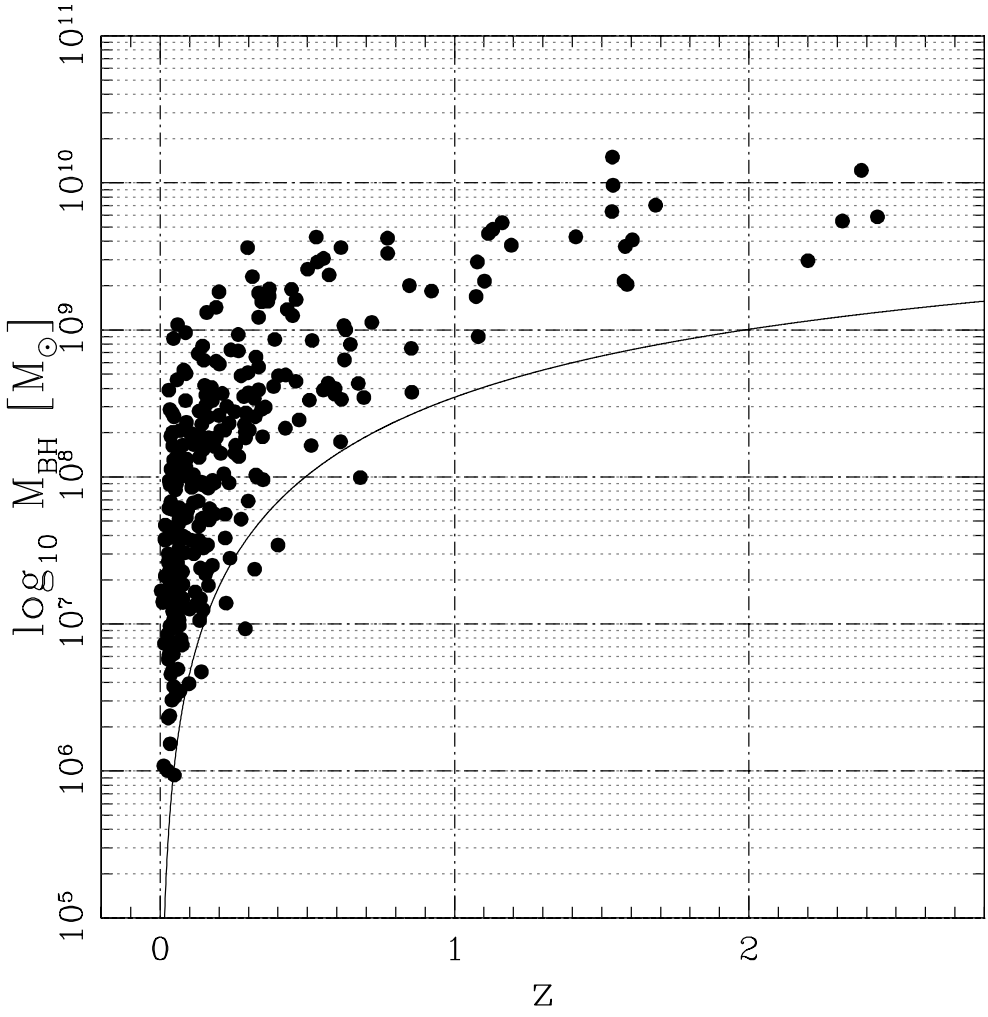}
\caption[]{Distribution of \mbh\ estimates as a function of $z$\
using uncorrected FWHM(\hbbc) values. Intermediate $z$ sources
observed with ISAAC are at $z \ga 0.8$. The solid curve estimates
the minimum detectable \mbh\ for our sample -- see
text.}\label{fig:massz}
\end{figure*}

\begin{figure*} \includegraphics[width=10.6cm, height=10.6cm,  angle=0]{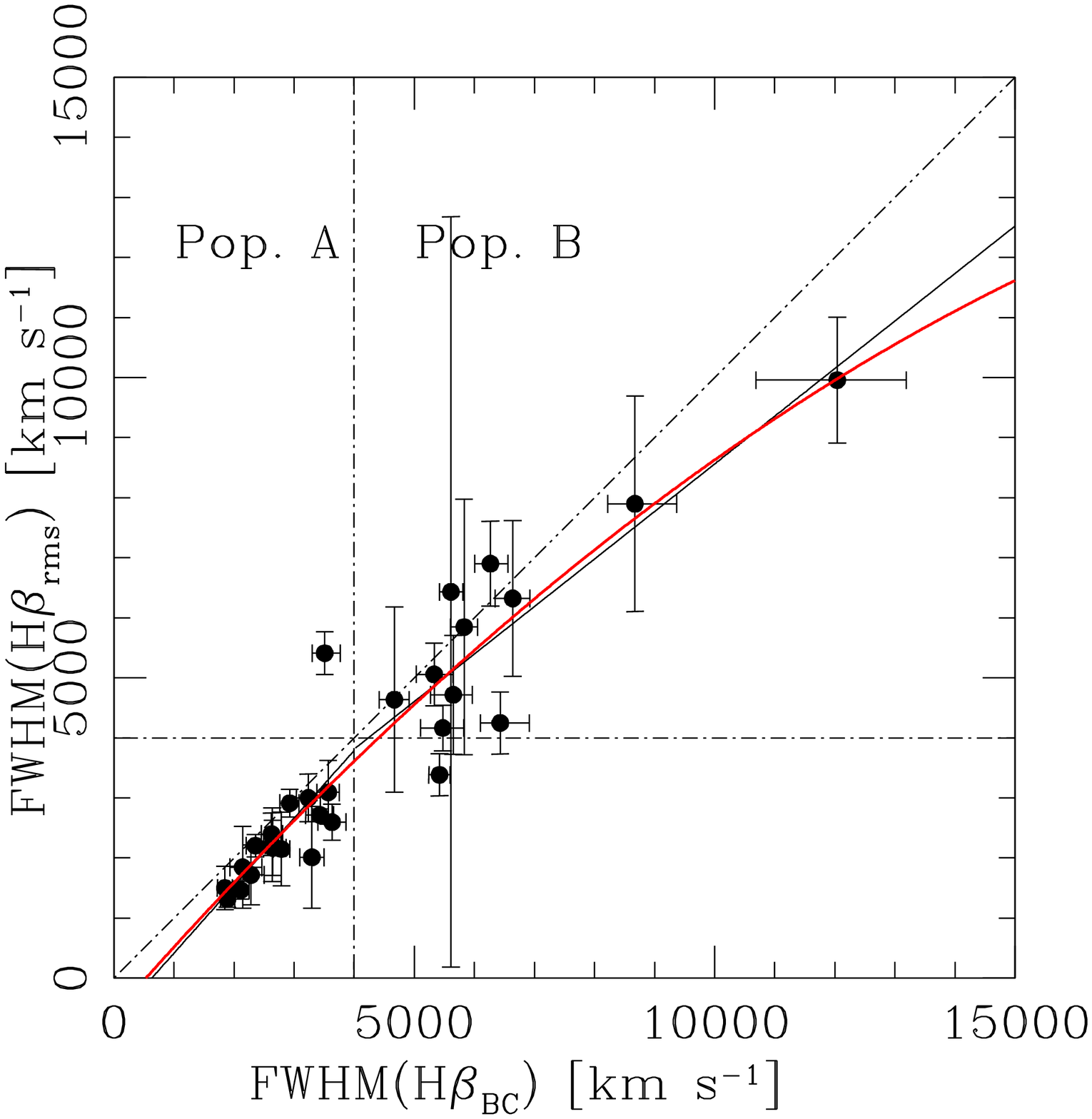}\\
\caption[]{FWHM  measured from the variable part of the \hbbc\
profile for low-$z$\ sources with reverberation data
\citep{petersonetal04} versus FWHM(\hbbc) (\kms) measures from
\citet{marzianietal03c}. Solid straight lines show best fits for
FWHM(\hbbc)$\le$ 4000 \kms (Pop. A) and for sources with broader
lines(Pop. B).  The solid curve shows a weighted least-square fit
of all data with a second order polynomial. The diagonal
dot-dashed line indicates the location of equal ordinate and
abscissa values.} \label{fig:peterson}
\end{figure*}

LILs like \mgii\ and \feii\ may yield more reliable results than
\hbbc. \hi\ Balmer line emission can be substantial from gas in a
variety of physical conditions. In Pop. B sources, a very broad
component may increase the FWHM of the integrated \hbbc\ profile
\citep[also mimicked by low $S/N$\
data;][]{mcintoshetal99,shemmeretal04}. This very broad component
may be optically thin to the ionizing continuum, and therefore
non-responsive to continuum changes. The \hbbc\ profile of Pop. A
sources may be affected by a high-ionization component mentioned
earlier.  \feii\ is thought to be emitted in a region very
optically thick to Lyman continuum which is probably photoionized
\citep{vestergaardpeterson05,wangetal05}, as the part of \hbbc\
responding to continuum changes should be. It is not surprising
that the reverberating part of \hbbc\ and \feii\ provides width
estimates which are consistent, since they are expected to measure
the width of a similar sub-region within the BLR. Similar
considerations apply to \mgii\ \citep{willsetal85}\ since \mgii\
should be mainly emitted in the same zone as \feii.
\citet{mcluredunlop04} present virial \mbh\ estimates for
$\approx$13000 quasars in the redshift interval $ 0.1 \la z \la
2.1 $ based on spectra from the SDSS first data release. The mean
\mbh\ increases with increasing redshift basically as shown in
Fig. \ref{fig:massz}. The mass values found by them are also
consistent with a limiting \mbh\ around $3 \cdot 10^9$ \msol, with
large scatter. They use FWHM(\hbbc) or FWHM(\mgii) and find a
consistency between the most massive at $ z \approx 2 $\ and those
at $ z \approx 0 $. Given measurement difficulties, and doubts
about the virial assumption for most other lines, corrected
measures for \hbbc, \feii, and \mgii\ may offer the best hope for
reliable \mbh\ and \lledd\ estimates out to $z \approx$ 2.5.

\section{Conclusions}

Nine intermediate $z$\ VLT/ISAAC spectra with high resolution and
$S/N$\ supplement an earlier sample of 17 sources. Emission line
measurements on \hbbc\ and \feii\ presented in this paper
strengthen the conclusion of \citet{sulenticetal04} that
luminosity effects are weak or absent in the low-ionization lines
of AGNs. Results on the \hbbc\ profile are consistent with the
population A-B hypothesis and Eigenvector 1 parameter space
concept developed for low $z$\ AGNs. We computed virial masses and
Eddington ratios for the 25 intermediate-$z$ objects plus about
280 lower $z$\ sources, using the same emission line over the
entire redshift range for the first time.

We also have  how the distributions of \mbh\ and \lledd\ vs. $z$\
are shaped by selection effects intrinsic to any flux-limited
survey at the low \mbh\ end. At the high \mbh\ end, masses
exceeding a few 10$^9$ \msol\ may be rare if corrections for
non-virial broadening and statistical errors are taken into
account. This suggestion is based on just 25 objects distributed
over the entire redshift range $0.9 \la z \la 2.5$. Confirmation
from a larger sample of intermediate-$z$\ observations is needed.


\newpage

\begin{acknowledgements}
We thank a referee for many useful suggestions and the tenacity to
ensure that they were taken into account. We also thank Lutz
Wisotzki for providing us with the HE optical spectra. D. D-H, and
JS acknowledge financial support from grant IN100703 from PAPIIT,
DGAPA, UNAM.
\end{acknowledgements}


\begin{table*}
\begin{center}
\caption{Basic Properties of Sources and Log of Observations \label{tab:obs}}
    \begin{tabular}{lllllllllllll}
    \hline  \hline
    \noalign{\smallskip}
     Object name & \multicolumn{1}{c}{$\rm m_B^{\mathrm{a}}$}
     &\multicolumn{1}{c}{$z^{\mathrm{b}}$} &
     \multicolumn{1}{c}{Line$^{\mathrm{c}}$} &\multicolumn{1}{c}{$M_{\mathrm B}^
     {\mathrm{d}}$}  & $\log \rm R_K^{\mathrm{e}}$ & Date$^{\mathrm{f}}$
     &
\multicolumn{1}{c}{Band$^{\mathrm{g}}$} & \multicolumn{1}{c}{DIT$^{\mathrm{h}}$}
     & \multicolumn{1}{c}{N$_{\rm exp}^{\mathrm{i}}$} & \multicolumn{1}{c}{Airmass$^{\mathrm{j}}$} &
     \multicolumn{1}{c}{S/N$^{\mathrm{k}}$} \\
\multicolumn{1}{c}{(1)}      & \multicolumn{1}{c}{(2)}         & \multicolumn{1}{c}{(3)}
     & \multicolumn{1}{c}{(4)}    & \multicolumn{1}{c}{(5)} & \multicolumn{1}{c}{(6)}  & \multicolumn{1}{c}{(7)}  & \multicolumn{1}{c}{(8)} & \multicolumn{1}{c}{(9)} & \multicolumn{1}{c}{(10)} & \multicolumn{1}{c}{(11)} & \multicolumn{1}{c}{(12)}\\
     \hline
     \noalign{\smallskip}
HE0507$-$3236&  17.36 &  1.5770 (7) & 1   &  $-$27.6 &$\la$ 0.51   &2003-10-22  & J    & 180   & 18  & 1.23-1.09 & 25--20\\
HE0512$-$3329&  17.03 &  1.5873 (7) & 1   &  $-$28.0 &$\la$ 0.38   &2003-10-22  & J    & 180   & 16  & 1.07-1.02 & 20--15\\
HE0926$-$0201&  16.23 &  1.6824 (7) & 1   &  $-$29.0 &$\la$ -0.33   &2004-02-11  & J    & 180   & 12  & 1.11-1.17 & 30--10\\
HE0946$-$0500&  16.24 &  1.1013 (7) & 1   &  $-$28.0 &$\la$ 0.03   &2004-02-11  & sZ   & 180   & 12  & 1.17-1.27 & 30\\
HE1003$+$0149&  16.45 &  1.0809 (7) & 1,2 &  $-$27.7 &      0.20   &2004-02-13  & sZ   & 180   & 12  & 1.19-1.14 & 20\\
HE1017$-$0009&  16.69 &  1.1295 (7) & 1   &  $-$27.6 &      0.87   &2004-02-28  & sZ   & 180   & 12  & 1.97-1.62 & 20--10\\
HE1104$-$1805&  16.45 &  2.3192 (7) & 1,2 &  $-$29.6 &$\la$ 0.19   &2004-02-11  & sH   & 180   & 20  & 1.14-1.35 & 20\\
HE1249$-$0648&  16.72 &  1.1940 (7) & 1   &  $-$27.7 &$\la$ 0.23   &2004-02-05  & sZ   & 180   & 16  & 1.04-1.06 & 30--15\\
HE1258$-$0823&  16.26 &  1.1632 (7) & 1   &  $-$28.1 &$\la$ 0.04   &2004-03-29  & sZ   & 180   & 17  & 1.06-1.16 & 30\\
\noalign{\smallskip} \hline \hline
\end{tabular}
\end{center}
\begin{list}{}{}
\item[$^{\mathrm{a}}$] Apparent B magnitude corrected because of Galactic absorption. \item[$^{\mathrm{b}}$] Redshift, with uncertainty in parenthesis.
\item[$^{\mathrm{c}}$] Lines used for redshift calculations: 1: \hb, 2: \oiii. \item[$^{\mathrm{d}}$] Absolute B
magnitude, computed for $H_0$=70 \kms Mpc$^{-1}$,  $\Omega_{\mathrm M} = 0.3$,  $\Omega_{\Lambda} = 0.7$, and $k$-correction spectral index $a$=0.6.
\item[$^{\mathrm{e}}$] Decimal logarithm of the specific flux ratio at 6cm and 4400 \AA\ (effective wavelength of  the
B band). Upper limits are from the NVSS ($\approx$ 2.5 mJy), and would place all undetected sources in the RQ domain.
\item[$^{\mathrm{f}}$] Date refers to time at start of exposure. \item[$^{\mathrm{g}}$] Photometric band.
\item[$^{\mathrm{h}}$] Detector Integration Time (DIT) of ISAAC, in seconds. \item[$^{\mathrm{i}}$] Number of
exposures with single exposure time equal to DIT. \item[$^{\mathrm{j}}$] Airmass at start and end of exposure.
\item[$^{\mathrm{k}}$] $S/N$\ at continuum level in the proximity of \hb. Two values are reported in case of different
$S/N$\  on the blue and red side of \hb\ (blue side first). The $S/N$\ value is with N estimated at a $2 \sigma$\
confidence level i.e., 2 times the rms.
\end{list}
\end{table*}

\begin{table*}
\begin{center}
\caption{Measurements of Fluxes, Equivalent Widths and FWHM of Strongest Lines} \label{tab:meas}
    \begin{tabular}{lcccc}
    \hline  \hline
    \noalign{\smallskip}
     Object name
     & \multicolumn{1}{c}{F(\hbbc)$^{\mathrm{a}}$}
     & \multicolumn{1}{c}{W(\hbbc)$^{\mathrm{b}}$}
     &\multicolumn{1}{c}{W(\feiiq)$^{\mathrm{c}}$}
     & \multicolumn{1}{c}{FWHM(\feiiq)$^{\mathrm{d}}$}
    \\
\multicolumn{1}{c}{(1)}      & \multicolumn{1}{c}{(2)}         & \multicolumn{1}{c}{(3)}
     & \multicolumn{1}{c}{(4)}    & \multicolumn{1}{c}{(5)} 
       \\
     \hline
     \noalign{\smallskip}
HE0507$-$3236 & 12.5 $\pm$ 3.0        &  62 $\pm$20             & 17 $\pm$ 3        & 4100 $^{+1600}_{-900}$ \\
HE0512$-$3329 & 20.0 $^{+5.0}_{-3.0}$  &  75 $^{+30}_{-10}$      & 47 $\pm$ 10       & 2400 $^{+1300}_{-500}$ \\
HE0926$-$0201 & 26.0 $\pm$ 4.0        &  72 $\pm$ 15            & 20 $\pm$ 2        & 3400 $\pm$ 1200        \\
HE0946$-$0500 & 17.0 $^{+3.0}_{-2.0}$  &  59 $^{+20}_{-10}$      & 19 $^{+5}_{-3}$   & 2900 $\pm$ 1000        \\
HE1003$+$0149 & 7.0 $\pm$ 1.0        &  48 $\pm$ 10            & 67 $\pm$ 12       & 2100 $\pm$ 1400        \\
HE1017$-$0009 & 10.5 $\pm$ 1.0  &  60 $^{+15}_{-10}$      & 11 $^{+5}_{-3}$   & 3600 $^{+2100}_{-1300}$\\
HE1104$-$1805 & 20.0 $\pm$ 4.0        &  72 $\pm$ 15            & 39 $^{+7}_{-5}$   & 4100 $\pm$ 1700        \\
HE1249$-$0648 & 15.5 $\pm$ 2.0        &  63 $\pm$ 15            & 31 $^{+5}_{-2}$   & 1400 $^{+1000}_{-0}$   \\
HE1258$-$0823 & 24.0 $\pm$ 3.0        &  44 $\pm$ 10            & 20 $\pm$ 4        & 1400 $^{+1000}_{-400}$ \\
\noalign{\smallskip} \hline \hline
\end{tabular}
\end{center}
\end{table*}

\begin{table*}
\begin{center}
    \begin{tabular}{lcccc}
    \hline  \hline
    \noalign{\smallskip}
     Object name
     &\multicolumn{1}{c}{F(\hbnc)$^{\mathrm{e}}$}
     &\multicolumn{1}{c}{W(\hbnc)$^{\mathrm{f}}$}
     & \multicolumn{1}{c}{W({\sc[Oiii]$\lambda4959$})$^{\mathrm{f}}$}
     & \multicolumn{1}{c}{W(\oiii)$^{\mathrm{f}}$}
    \\
\multicolumn{1}{c}{(6)}      
     & \multicolumn{1}{c}{(7)}    & \multicolumn{1}{c}{(8)} &
     \multicolumn{1}{c}{(9)} & \multicolumn{1}{c}{(10)}
       \\
     \hline
     \noalign{\smallskip}
HE0507$-$3236  & 7:        & 0.3:       & 2.7 $\pm$ 0.5      & 8.4 $\pm$  0.6 \\
HE0512$-$3329  & 3:        & 0.1:       & 2.5 $\pm$ 0.5      & 6.5 $\pm$  1.0 \\
HE0926$-$0201  & 28:        & 0.8:       & 5.6 $\pm$ 0.5      &15.2 $\pm$  1.0 \\
HE0946$-$0500  & 14 $^{+9}_{-5}$     & 0.9         &1.1 $\pm$   0.5  & 3.4 $\pm$  0.6 \\
HE1003$+$0149  & 15 $^{+18}_{-5}$     & 1.0       &0.6 $\pm$ 0.5    & 1.7 $\pm$  0.5 \\
HE1017$-$0009  & 15 $^{+10}_{-5}$     & 0.8       &1.4 $\pm$ 0.5    & 3.8 $\pm$  0.6 \\
HE1104$-$1805  & 60 $ \pm$ 18           & 2.0 $\pm$ 0.5      &3.9 $\pm$ 0.5    &11.4 $\pm$  0.5 \\
HE1249$-$0648  & 18 $ \pm$ 5           & 0.7 $\pm$ 0.3      & ...             & ...            \\
HE1258$-$0823  & 9:                   & 0.8 $\pm$ 0.3      & ...             & ...            \\
\noalign{\smallskip} \hline \hline
\end{tabular}
\end{center}
\begin{list}{}{}
\item[$^{\mathrm{a}}$] Rest frame flux of \hbbc\ in units of 10$^{-13}$\ \ergss\ cm$^{-2}$  and  $\pm 2\sigma$
confidence level uncertainty. \item[$^{\mathrm{b}}$] Rest frame equivalent width of \hbbc\ in \AA\ $\pm 2\sigma$
confidence level uncertainty. \item[$^{\mathrm{c}}$] Rest frame equivalent width of the \feiiq\ blend in \AA\ $\pm
2\sigma$ confidence level uncertainty. \item[$^{\mathrm{d}}$] FWHM of lines in the \feiiq\ blend and uncertainty at
2$\sigma$, in \kms. See text for details. \item[$^{\mathrm{e}}$] Rest frame flux of \hbnc\ in units of 10$^{-16}$\
\ergss\ cm$^{-2}$  and  $\pm 2\sigma$ confidence level uncertainty.  Colons indicate highly uncertain values.
\item[$^{\mathrm{f}}$] Rest frame equivalent width of \hbnc, {\sc [Oiii]}\l 4959, and \oiii\ in \AA, with uncertainty
at 2$\sigma$. Colons indicate highly uncertain values.
\end{list}
\end{table*}

\begin{table*}
\begin{center}
\caption{\hbbc\ Line Profile Measurements} \label{tab:profs}
    \begin{tabular}{lllccrrrrrr}
    \hline  \hline
    \noalign{\smallskip}
     \multicolumn{1}{c}{Source}
     & \multicolumn{1}{c}{FWZI$^{\mathrm{a}}$}
     & \multicolumn{1}{c}{$\Delta^{\mathrm{a,b}}$}
     & \multicolumn{1}{c}{FWHM$^{\mathrm{a}}$}
     & \multicolumn{1}{c}{$\Delta^{\mathrm{a,b}}$}
     & \multicolumn{1}{c}{A.I.$^{\mathrm{c}}$}
     & \multicolumn{1}{c}{$\Delta^{\mathrm{b}}$}
     & \multicolumn{1}{c}{Kurt.$^{\mathrm{d}}$}
     & \multicolumn{1}{c}{$\Delta^{\mathrm{b}}$}
     & \multicolumn{1}{c}{c(0/4)}
     & \multicolumn{1}{c}{$\Delta^{\mathrm{a,b}}$}
     \\
\multicolumn{1}{c}{(1)}      & \multicolumn{1}{c}{(2)}         & \multicolumn{1}{c}{(3)}
     & \multicolumn{1}{c}{(4)}    & \multicolumn{1}{c}{(5)} & \multicolumn{1}{c}{(6)} & \multicolumn{1}{c}{(7)}
     & \multicolumn{1}{c}{(8)}   & \multicolumn{1}{c}{(9)}  & \multicolumn{1}{c}{(10)}
      & \multicolumn{1}{c}{(11)} \\    \hline
     \noalign{\smallskip}
HE0507$-$3236 &16000 &2100 &3500  &300 &$-$0.15 &0.10                 & 0.31  & 0.06  &$-$500     &1100\\
HE0512$-$3329 &16000 &3700 &3100  &300 &$-$0.16 &0.07                 & 0.31  & 0.04  &$-$100     &1900\\
HE0926$-$0201 &17000 &1400 &5100  &500 &   0.21 &$^{+0.07}_{-0.19}$   & 0.27  & 0.06  &   1200    &700\\
HE0946$-$0500 &20000 &2800 &3600  &250 &$-$0.06 &0.09                 & 0.37  & 0.05  &$-$600     &1400\\
HE1003$+$0149 &9000  &2100 &2900  &280 &$-$0.06 &$^{+0.07}_{-0.11}$   & 0.33  & 0.05  &$-$400     &1000\\
HE1017$-$0009 &16000 &3800 &6200  &290 &$-$0.04 &$^{+0.08}_{-0.05}$   & 0.41  & 0.04  &   200     &1900\\
HE1104$-$1805 &15000 &1500 &4300  &260 &   0.11 &0.09                 & 0.38  & 0.05  &   900    &800\\
HE1249$-$0648 &21000 &1000 &4900  &500 &   0.13 &$^{+0.11}_{-0.07}$   & 0.31  & 0.05  &   1400    &500\\
HE1258$-$0823 &17000 &2600 &4400  &500 &$-$0.08 &$^{+0.20}_{-0.08}$   & 0.23  & 0.06  &$-$700    &1300\\
\noalign{\smallskip} \hline \hline
\end{tabular}
\end{center}
\end{table*}

\begin{table*}
\begin{center}
    \begin{tabular}{lrrrrrrrr}
    \hline  \hline
    \noalign{\smallskip}
     \multicolumn{1}{c}{Source}
    & \multicolumn{1}{c}{c(1/4)}
     & \multicolumn{1}{c}{$\Delta^{\mathrm{a,b}}$}
     & \multicolumn{1}{c}{c(1/2)$^{\mathrm{a}}$}
     & \multicolumn{1}{c}{$\Delta^{\mathrm{a,b}}$}
     & \multicolumn{1}{c}{c(3/4)$^{\mathrm{a}}$}
     & \multicolumn{1}{c}{$\Delta^{\mathrm{a,b}}$}
     & \multicolumn{1}{c}{c(0.9)$^{\mathrm{a}}$}
     & \multicolumn{1}{c}{$\Delta^{\mathrm{a,b}}$}
     \\
\multicolumn{1}{c}{(12)}      
     & \multicolumn{1}{c}{(13)}
     & \multicolumn{1}{c}{(14)}
     & \multicolumn{1}{c}{(15)}
     & \multicolumn{1}{c}{(16)}  & \multicolumn{1}{c}{(17)}
      & \multicolumn{1}{c}{(18)}  & \multicolumn{1}{c}{(19)}  &
      \multicolumn{1}{c}{(20)}
    \\    \hline
     \noalign{\smallskip}
HE0507$-$3236  &$-$300       &300      &       50   &150  &    100  &90   &   90    &50\\
HE0512$-$3329  &$-$500       &200      &    $-$180    &150  &$-$80  &80   &$-$30    &50\\
HE0926$-$0201  &   800       &$_{-1100}^{+400}$&    $-$150    &250  &$-$160 &130  &$-$110   &80\\
HE0946$-$0500  &$-$300       &200      &    $-$320    &120  &$-$160 &100  &$-$70    &70\\
HE1003$+$0149  &$-$100       &200      &    $-$50     &130  &$-$20  &70   &   10    &50\\
HE1017$-$0009  &$-$900       &300      &    $-$610    &130  &$-$350 &140  &$-$360   &120\\
HE1104$-$1805  &   200       &300      &    $-$180     &130  &$-$190  &120  &$-$190    &80\\
HE1249$-$0648  &   400       &400      &    $-$70     &240  &$-$60  &130  &$-$110   &80\\
HE1258$-$0823  &$-$400       &$_{-300}^{+700}$ &    $-$330    &220  &$-$30  &110  &$-$40    &50\\

\noalign{\smallskip} \hline \hline
\end{tabular}
\end{center}
\begin{list}{}{}
\item[$^{\mathrm{a}}$] In units of \kms. \item[$^{\mathrm{b}}$] 2$\sigma$ confidence level uncertainty.
\item[$^{\mathrm{c}}$] Asymmetry index defined as in Marziani et al.\ (1996). \item[$^{\mathrm{d}}$] Kurtosis
parameter as in Marziani et al.\ (1996).
\end{list}
\end{table*}
\vfill
\pagebreak
\newpage

\begin{figure*}
\includegraphics[width=11.4cm, height=11.4cm, angle=0]{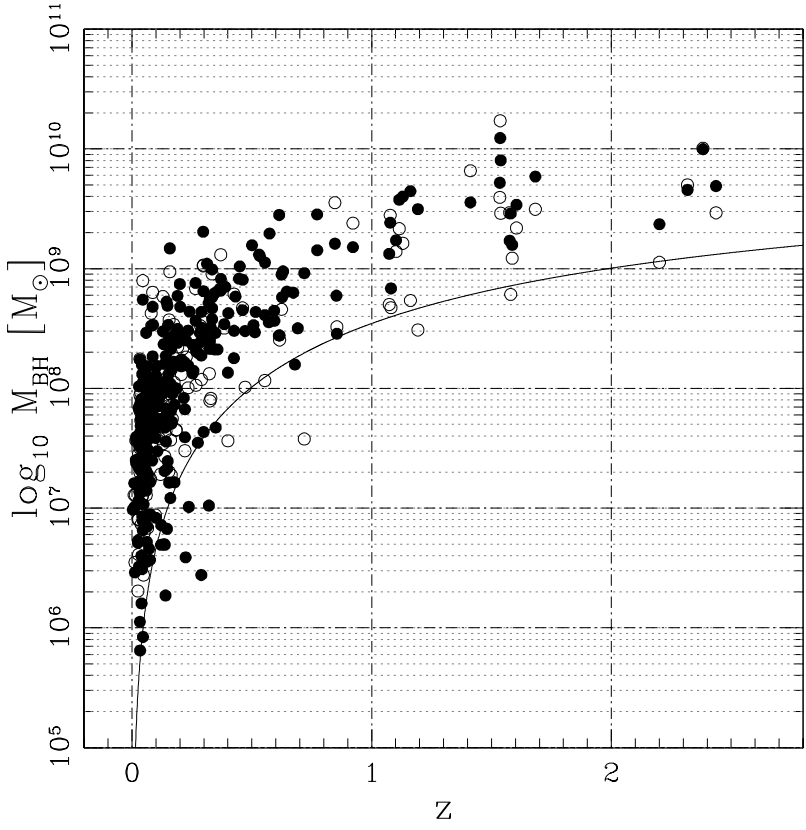}\\
\caption[]{\mbh\ computed from corrected
FWHM$^{\mathrm{corr}}$(\hbbc) as a function of $z$\ (filled
circles). \mbh\ estimates computed from FWHM(\feiiq) measures are
shown as open symbols. Symbol types and solid curve are  as
described  for Fig. \ref{fig:massz}. }\label{fig:masszpet}
\end{figure*}

\begin{figure*}
\includegraphics[width=9cm, height=9cm, angle=0]{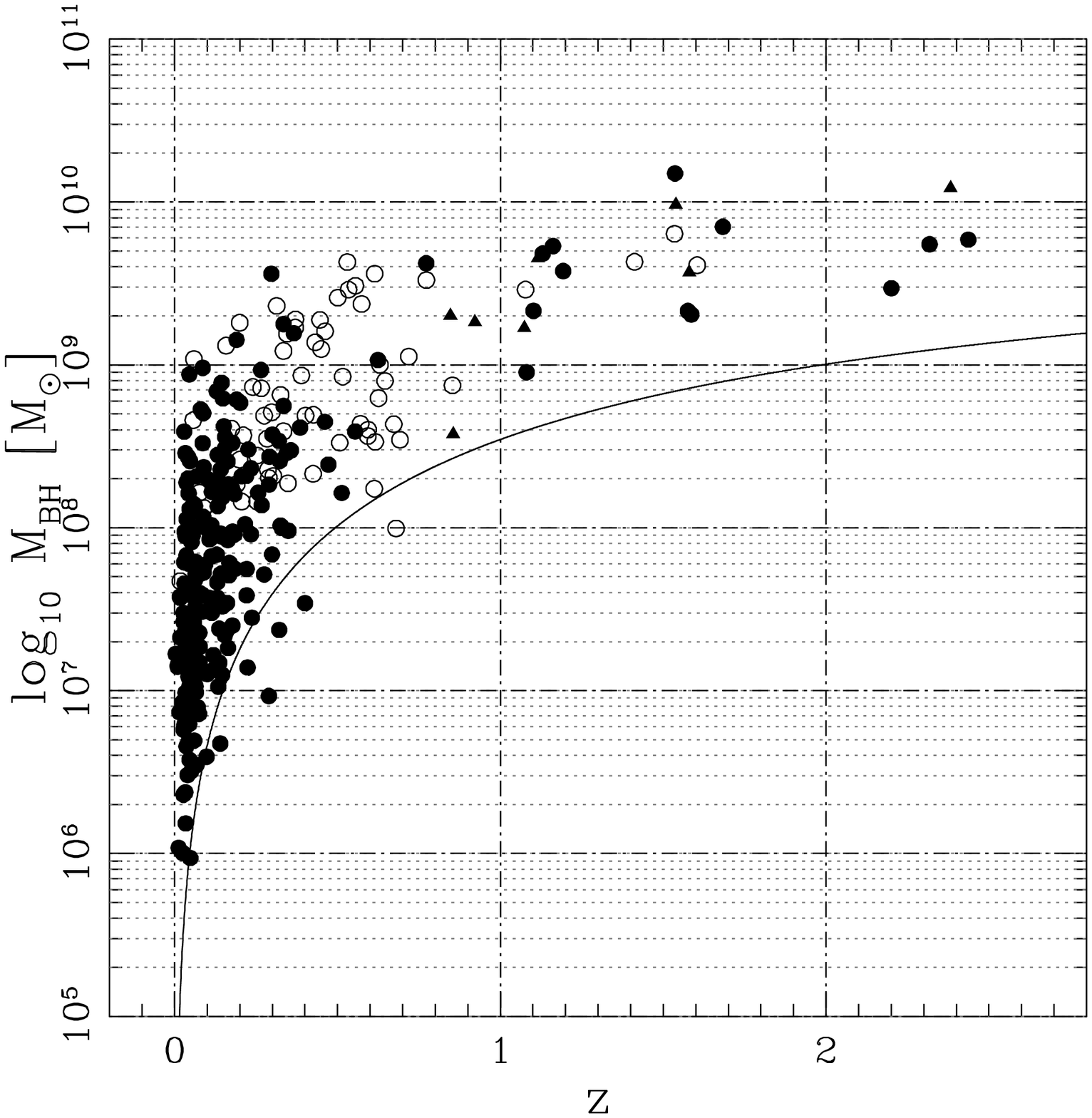}
\includegraphics[width=9cm, height=9cm, angle=0]{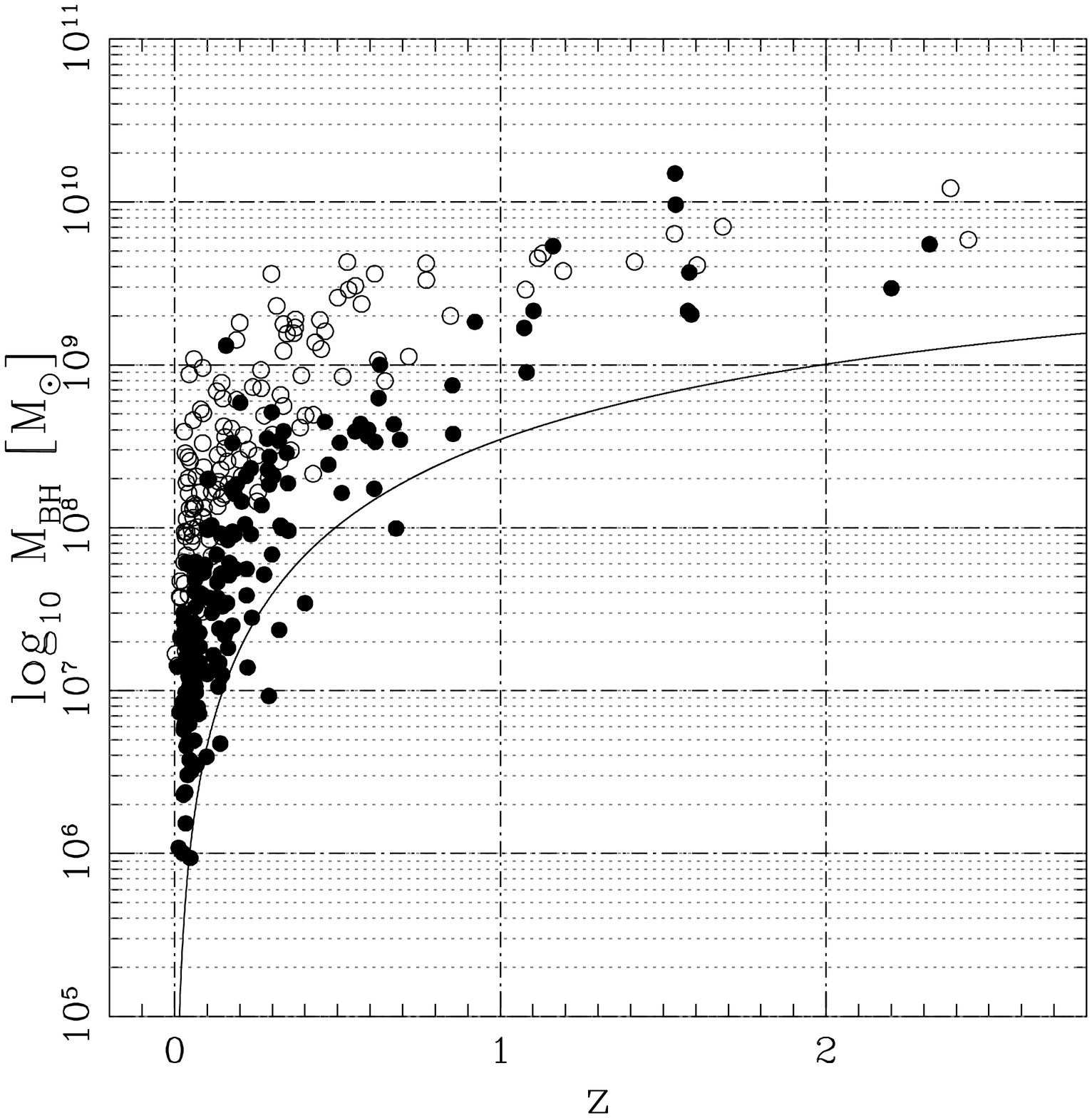}
\caption[]{\mbh\ computations using  same samples as Fig. \ref{fig:massz}, where RQ and RL sources are indicated by
filled and open symbols respectively (upper panel). Triangles indicate sources for which radio data are insufficient
to compute a meaningful \rk. The lower panel distinguishes \mbh\ for Pop. A (filled symbols) and Pop. B (open
symbols). The solid curve is as described for Fig. \ref{fig:massz}. } \label{fig:masszrlq}
\end{figure*}

  \begin{figure*}
\includegraphics[width=11.0cm, height=11.0cm,
  angle=0]{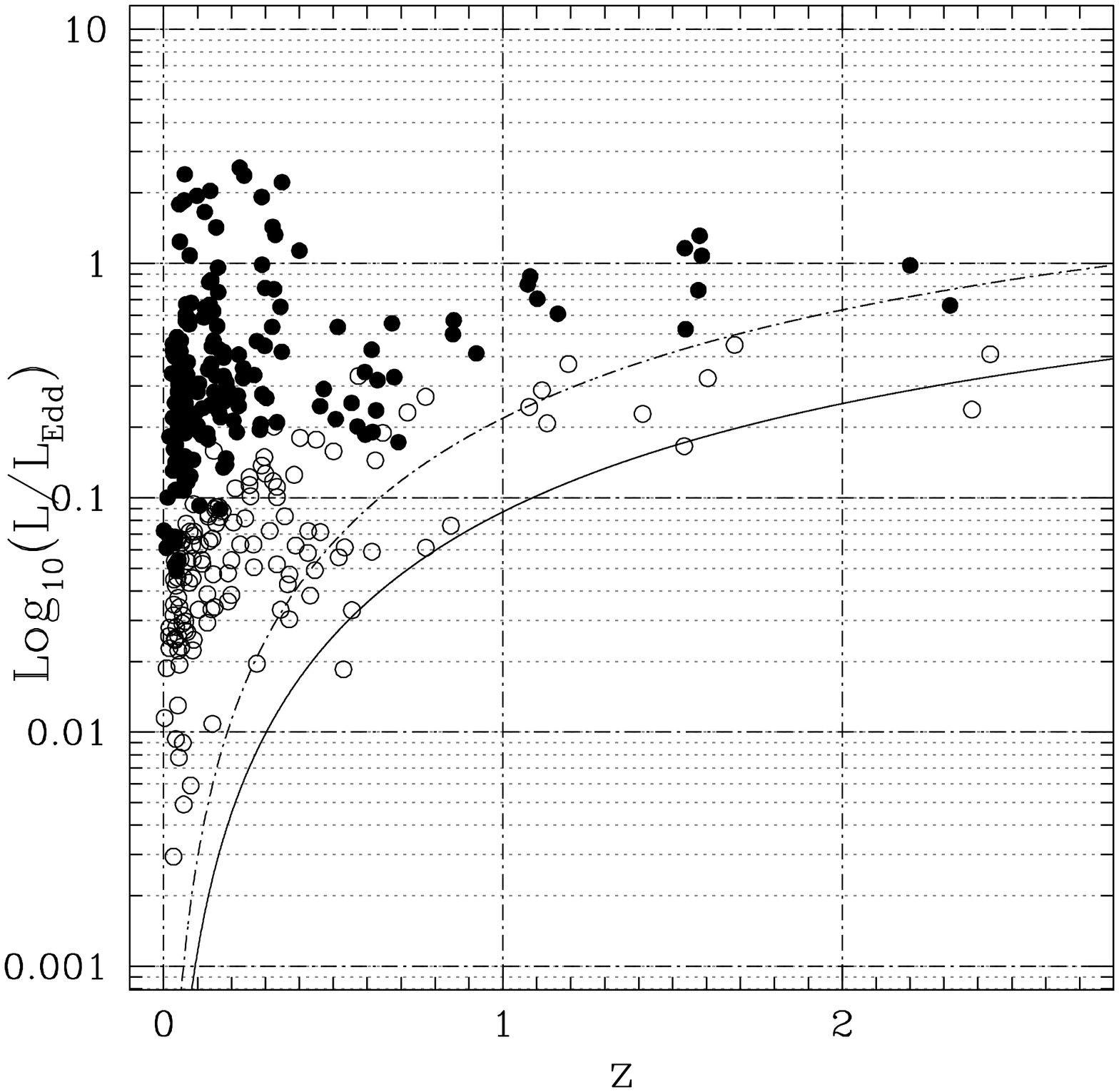}
\caption[]{Distribution of \lledd\ with $z$ following Fig.
\ref{fig:massz} with  $\lambda L_\lambda$ values derived from the
5100 \AA\ rest-frame flux and with $\alpha = 0.67$. E1 Populations
A and B are indicated by filled and open symbols respectively. The
solid line traces the minimum \lledd\ for a fixed mass value of $
4 \cdot 10^{9}$\ \msol\ to observe a quasar above the limiting
magnitude of the HE quasar survey $m_{\mathrm B } \approx 17.5$.
The dot-dashed lines indicate the same limit but assuming limiting
magnitude $m_{\mathrm B } \approx 16.5$. } \label{fig:lmz}
\end{figure*}
\vfill\eject

\newpage

\appendix

\section{{Notes on Individual Sources} \label{individual}}

\subsection{HE0512-3329}


HE 0512--3329 was discovered as a probable gravitationally lensed quasar with the Space Telescope Imaging Spectrograph
(STIS). It is a doubly imaged QSO with a source redshift of $ z = 1.58 $\ and an image separation of 0''.644. The flux
ratio $ \frac{A}{B} $ of the lensed images shows a strong dependence on wavelength. In the R and I bands, A is
brighter than B by about 0.45 mag while the two are almost equal in the band B. For smaller wavelengths, especially
close to the limit near 2000  \AA, B becomes much brighter than A by about 1.3 mag. A natural explanation for this
effect is differential reddening caused by different extinction effects in the two lines of sight \citep{greggetal00}.
Microlensing by stars and other compact object in the lensing galaxy also plays a role in this source
\citep{wucknitzetal03}. Unfortunately, the small separation does not allow to us to distinguish component A and B on
the ISAAC spectrum. The HE 0512--3329 acquisition image is compatible with an unresolved source. Considering the flux
ratios in the R and I band, it is reasonable to assume that our spectrum is dominated by the A component.

\subsection{HE1104-1805}

HE 1104-1805 is a double-image lensed quasar discovered by
\citet{wisotzkietal93}. The image separation is $ \Delta \theta =
3''.19 $, the source redshift is $ z_s = 2.319 $, and the lens
redshift is $ z_l = 0.729 $. \citet{wisotzkietal95} reported that
the continuum flux in both images is highly variable but that the
line fluxes do not change, as expected if  microlensing is
operating. On the acquisition image of HE 1104 there is a second
source at ~3.5" but it is completely off-slit.

\vfill\newpage

\section{Minimum \mbh\ and \lledd\ as a Function of $z$ \label{appendix}}

The absolute B magnitude \Mb\ can be related to the specific luminosity at 5100 \AA\ assuming an average spectral
shape between 4400 (effective wavelength of B band) and 5100 \AA. If the spectral shape is described by a power-law
($f_\nu \propto \nu^{\mathrm -b}$) with $b = 0.3$\ \citep{marzianietal03b}:

\begin{equation}
\log [\lambda L_\lambda (\lambda = 5100 {\mathrm \AA})] = -0.4 M_{\rm B} + 35.497. \label{eq:m01}
\end{equation}

If we further assume that $L = 10\lambda \cdot L_\lambda$, with $\lambda$=5100 \AA\ i.e., a bolometric correction
factor 10,  we can write:

\begin{equation}
\log [\lambda L_\lambda (\lambda = 5100 {\mathrm \AA})] = \frac{L} {10 \cdot L_{\mathrm Edd}} \cdot 1.3 \cdot 10^{47}
M_{\mathrm{BH},9\odot}\label{eqn:bolcorr}
\end{equation}

where the black hole mass \mbh\ has been written in units of $10^9$ \msol. The absolute magnitude \Mb\ is:

\begin{equation}
M_{\mathrm B} = m_{\mathrm B} + 5 - 5 \log d_{\mathrm L} - 2.5 (a - 1) \log (1 + z)
\end{equation}

where the luminosity distance $d_{\mathrm L}$\ is in parsec, and the last term is the $K(z)$\ correction.

Assuming $\Omega_{\mathrm M} \neq\ 0, \Omega_{\Lambda} \neq\ 0$,\ and  $\Omega_{\mathrm k} = 0$, we have for the
co-moving distance:

\begin{equation}
 d_{\mathrm C} = \frac{c}{H_0} \int_0^z \frac{dz'}{\sqrt{\Omega_{\mathrm M} (1+z)^3 + \Omega_\Lambda}}
\end{equation}

This integral can be approximated with  residuals less than 3\% at all $z$\ (maximum error for $z \rightarrow 0$).

\begin{equation}
 d_{\mathrm C} \approx \frac{c}{H_0} \left[ (1.500 ) \left( 1 - e^{-\frac{z}{6.107 }} \right) +
0.996  \left( 1 - e^{-\frac{z}{1.266  }} \right) \right]
 \end{equation}

The luminosity distance is:
\begin{equation}
 d_{\mathrm L} = d_{\mathrm C} (1 + z)
\end{equation}
and we can use the new relationship to derive a working relationship for \mbh:

\begin{eqnarray*}
{M_{{\mathrm{BH}},9\odot}} = &    &  {3.85 \cdot 10^6 h^{-2}}
{\left( \frac{L}{L_{\mathrm Edd}}\right) }^{-1}  \cdot
10^{-0.4 m_{\mathrm B}} \cdot \\
& &  \cdot \left[ 1.5  \left( 1 - e^{-\frac{z}{6.107 }} \right) +
\left( 1 - e^{-\frac{z}{1.266}} \right) \right]^2 (1 + z)^{(1 -
\mathrm{a})}, \\ \label{eqn:massapp}
\end{eqnarray*}



where $h = H_0/75$. Curves shown in Figs. \ref{fig:massz}, \ref{fig:masszpet}, \ref{fig:masszrlq}, \ref{fig:lmz}
assume $a = 0.6$. Differences for $0.3 \simlt a \simlt 0.6$\ are minor. \vfill\newpage

%
%

\newpage

\end{document}